\documentclass[
  twocolumn,
  prb,
  showpacs,
  amsmath,
  amssymb,
  superscriptaddress,
  floatfix
]{revtex4}

\renewcommand{\cite}[1]{{[}\onlinecite{#1}{]}}

\usepackage{bm}
\usepackage{dsfont}
\usepackage{graphicx}
\usepackage{pifont}

\usepackage{color}

\newcommand{\s}{\sum\limits}

\newcommand{\pa}{\partial}

\newcommand{\be}{\begin{equation}}
\newcommand{\e}{\end{equation}}
\newcommand{\beml}{\begin{subequations}}
\newcommand{\eml}{\end{subequations}}
\newcommand{\beq}{\begin{eqnarray}}
\newcommand{\eq}{\end{eqnarray}}
\newcommand{\ba}{\begin{array}}
\newcommand{\ea}{\end{array}}
\newcommand{\bpm}{\begin{pmatrix}}
\newcommand{\epm}{\end{pmatrix}}
\newcommand{\bc}{\begin{cases}}
\newcommand{\ec}{\end{cases}}
\newcommand{\lt}{\left}
\newcommand{\rt}{\right}
\newcommand{\n}{\nonumber}

\newcommand{\ep}{\varepsilon}

\newcommand{\bb}{\boldsymbol}
\newcommand{\h}{^\dagger}

\DeclareMathOperator{\tr}{Tr}
\DeclareMathOperator{\diag}{diag}

\begin{document}

\title{Quantum Hall criticality and localization in graphene with short-range impurities at the Dirac point}

\author{S.~Gattenl\"ohner}
\affiliation{
 School of Engineering \& Physical Sciences, Heriot-Watt University,
 Edinburgh EH14 4AS, UK
}

\author{W.-R.~Hannes}
\affiliation{
 School of Engineering \& Physical Sciences, Heriot-Watt University,
 Edinburgh EH14 4AS, UK
}

\author{P.~M.~Ostrovsky}
\affiliation{Max-Planck-Institut f\"ur
Festk\"orperforschung, Heisenbergstr. 1, 70569, Stuttgart, Germany}
\affiliation{
 L.~D.~Landau Institute for Theoretical Physics RAS,
 119334 Moscow, Russia
}

\author{I.~V.~Gornyi}
\affiliation{
 Institut f\"ur Nanotechnologie, Karlsruhe Institute of Technology,
 76021 Karlsruhe, Germany
}
\affiliation{
 A.F.~Ioffe Physico-Technical Institute,
 194021 St.~Petersburg, Russia.
}

\author{A.~D.~Mirlin}
\affiliation{
 Institut f\"ur Nanotechnologie, Karlsruhe Institute of Technology,
 76021 Karlsruhe, Germany
}
\affiliation{Institut f{\"u}r Theorie der kondensierten Materie and DFG Center for Functional Nanostructures,
Karlsruhe Institute of Technology, 76128 Karlsruhe, Germany
}
\affiliation{Petersburg Nuclear Physics Institute, 188300 St. Petersburg, Russia
}

\author{M.~Titov}
\affiliation{
Radboud University Nijmegen, Institute for Molecules and Materials, NL-6525 AJ Nijmegen, The Netherlands}

\begin{abstract}
We explore the longitudinal conductivity of graphene at the Dirac point in a
strong magnetic field with two types of short-range scatterers: adatoms that mix
the valleys and ``scalar'' impurities that do not mix them. A scattering theory
for the Dirac equation is employed to express the conductance of a graphene
sample as a function of impurity coordinates; 
an averaging over impurity positions is then performed numerically.  
The conductivity $\sigma$ is equal to the ballistic value $4e^2/\pi h$ for each 
disorder realization provided the number of flux quanta
considerably exceeds the number of impurities. For weaker fields, the
conductivity in the presence of scalar impurities scales to the quantum-Hall
critical point with $\sigma \simeq 4 \times 0.4 e^2/h$ at half filling or to
zero away from half filling due to the onset of Anderson localization. For
adatoms, the localization behavior is obtained also at half filling due to
splitting of the critical energy by intervalley scattering. Our results reveal 
a complex scaling flow governed by fixed points of different symmetry classes: 
remarkably, all key manifestations of Anderson localization 
and criticality in two dimensions are observed numerically in a single setup.

\end{abstract}


\maketitle


The discovery of graphene has initiated an intense study of its electronic
properties~\cite{geim07,graphene-review}. 
From the fundamental point of view, the interest to graphene is largely
motivated by the quasirelativistic character of its spectrum: charge carriers in
graphene are two-dimensional (2D) massless Dirac fermions. This leads to a
variety of remarkable phenomena governed by the inherent topology of  Dirac fermions
as well as by their physics in the presence of various
types of disorder and interactions. 

Controllable functionalization of the graphene surface is possible by a variety of
tools, including hydrogenation \cite{ryu08,Elias09,burgess11,Katoch10},
fluorination \cite{robinson10}, adsorption of gas molecules
\cite{schedin07,Liu08}, ion irradiation \cite{fuhrer09,fuhrer11}, electron-beam
irradiation \cite{balandin}, and deposition of metallic islands \cite{Bouchiat}.
Adatoms, deposited molecules or islands, and defects engineered in this way
serve as strong short-range scatterers as has been also supported by a density
functional theory analysis \cite{Lichtenstein}. Furthermore, such scatterers
also exist in pristine graphene and may dominate its transport properties (in
particular, in suspended devices \cite{andrei08,bolotin}). It is thus important
to theoretically explore the transport in graphene with this kind  of disorder.
Away from the Dirac point the conductivity of such structures is sufficiently
well understood
\cite{ostrovsky06,Stauber07,novikov08,basko08,robinson08,pereira08,Yuan10}. 

Electronic transport properties of graphene near zero energy (Dirac point) are
particularly exciting. Remarkably, experiments  discovered
\cite{novoselov05b,kim05} that the conductivity of graphene at the Dirac point
is essentially independent of temperature in a broad range (from 300\,K down to
30\,mK) and has a value close to the conductance quantum $e^2/h$ (times four, which is the
total spin and valley degeneracy). This discovery attracted a great interest
because in conventional materials, once the conductivity is close to $e^2/h$, it
becomes strongly suppressed by Anderson localization with lowering temperature.
The above experiments, thus, indicate that the Dirac-point physics of disordered
graphene may be controlled by the vicinity of some quantum critical point. Indeed,
theoretical investigations have shown that for certain classes of disorder that
preserve some of symmetries of the clean Dirac Hamiltonian, the system avoids
Anderson localization \cite{ostrovsky06,ostrovsky07,evers08}.
Recent work \cite{ponomarenko11} demonstrated the feasibility of the
systematic experimental study of localization near the Dirac point.

A quantizing magnetic field yields a further remarkable twist to the fascinating
physics of graphene near the Dirac point. The Dirac character of the spectrum makes
graphene a unique example of a system where the quantum Hall (QH) effect can
be observed up to the room temperature \cite{novoselov07}. The zeroth Landau
level in graphene, with half of its edge branches being electron-like and
another half hole-like \cite{AQHE,Abanin07a}, has no analogue in semiconducting
2D electron systems with parabolic dispersion. Recent works on graphene in a
strong magnetic field show fractional QH effect as well as an insulating
behavior at the Dirac point, indicating a splitting of the critical energy in the
zeroth Landau level \cite{Ong08,Du09,Bolotin09,Kim12}.  

The goal of this paper is to study the Dirac-point conduction of graphene with
randomly positioned strong short-range impurities in a quantizing transverse
magnetic field. An analytical theory of transport in graphene developed in
Refs.~\cite{ostrovsky06,ostrovsky07,evers08,Ostrovsky08} maps the problem onto a
field theory---non-linear $\sigma$ model (NL$\sigma$M)---that is subsequently
analyzed by renormalization group (RG) means. The crucial role in this analysis
is played by the symmetry class and the topology of the NL$\sigma$M. These field
theories possess in 2D a rich family of non-trivial fixed points. Scaling flow
between these fixed points governs the evolution of conductivity with increasing
system size or decreasing temperature.  

It is important to stress, however, that the NL$\sigma$M is fully controllable
at conductivity $\sigma \gg e^2/h$, while the physics near zero energy and the
corresponding fixed points correspond to $\sigma \sim e^2/h$ (strong coupling
for the field theory). Thus, an extrapolation of the theory is required. While
this is a common ideology in condensed matter physics, one cannot a priori
exclude the possibility of a more complex behavior not anticipated from the weak
coupling expansion. Thus, a numerical modeling is of paramount importance to
probe the strong-coupling physics. In addition to verifying qualitative
predictions of the NL$\sigma$M, such a modeling should give values of
conductivity at fixed points and quantitatively characterize crossovers between
them. A number of works have studied transport and localization in
graphene with scatterers near the Dirac point
\cite{Yuan10,roche-localization,roche-vacancies,mayou} by numerical
analysis of the Kubo conductivity or of wave packet propagation. The results for
disorder formed by a low concentration of vacancies
\cite{Yuan10,roche-vacancies,mayou} disagree with expectations based
on NL$\sigma$M that the chiral symmetry inhibits localization.  

In this paper, we use the unfolded scattering theory
\cite{Titov10,Ostrovsky10,Schelter11} for the Dirac Hamiltonian with point-like
scatterers and extend it to incorporate magnetic fields. This allows
us to get an exact expression for the conductivity for a given configuration of
impurities. Subsequent av\-er\-aging over impurity positions is performed
numerically. This combination of analytical and numerical tools turns out to be
a very efficient way of evaluating the conductivity of graphene with rare
short-range scatterers.  

We consider two types of impurities: i) ``scalar'' impurities represented by a
smooth electrostatic potential that does not mix the graphene valleys and ii)
adatoms represented by an on-site impurity potential mixing the valleys.  Our
results yield a rich scaling flow controlled by a number of fixed points of
different symmetry classes, in qualitative agreement with predictions based on
the $\sigma$-model.  We also demonstrate that in the limit of high magnetic
field such that the number of flux quanta piercing the sample exceeds the number
of impurities, the conductivity returns to its ballistic value, $\sigma_{xx}
\approx 4e^2/\pi h$, for each disorder realization. 

Electronic properties of clean graphene are modeled by the Dirac
Hamiltonian, $H_{\bb{A}}= v\, \bb{\sigma} (\bb{p} - e \bb{A}/c)$, where
$\bb{\sigma}=(\sigma_x,\sigma_y)$, $\bb{p}$ is the 2D momentum
operator, $\bb{A}$ is the vector potential, and $v\approx 10^6$m/s is
the electron velocity. The disorder potential is
given by a superposition of individual impurity potentials. The distance
between impurities is assumed to be much larger than both the lattice constant
and the spatial range of an impurity. 

We consider a rectangular graphene sample with periodic
boundary conditions in $y$ direction ($0<y<W$) and open boundary conditions in
$x$ direction ($0<x<L$). The latter correspond to highly doped graphene leads
\cite{Tworzydlo06,Titov10}.  The transport properties of this setup are
described by a generating function
\cite{Schuessler09,Titov10,Ostrovsky10,Schelter11}, $\mathcal{F}(\phi)$,
whose derivatives with respect to the fictitious source field $\phi$ are related to the moments of the transmission distribution. In particular, Landauer's formula for the conductance $\mathrm{G}$ can be written as $\mathrm{G}=(4e^2/h) \partial^2
\mathcal{F}/\partial\phi^2|_{\phi=0}$.  Within the ``unfolded scattering theory''
\cite{Titov10,Ostrovsky10}, the impurity contribution to $\mathcal{F}(\phi)$ is expressed in terms of
Green functions $G_{\bb{A}}(\bb{r}_m,\bb{r}_n; \phi)$ of the clean system connecting impurity
positions and the matrix $\hat{T}=\diag (T_1,T_2,\dots, T_{N})$ constructed from
individual impurity T-matrices in the $s$-wave approximation at zero energy; see the Supplemental
Material \cite{EPAPS} for details.

A remarkable feature of the zero-energy state of clean graphene is the
existence of a non-unitary gauge transformation, making it possible to
gauge away the entire magnetic field \cite{Schuessler09}. This transformation
can be formulated as
\be
\label{trans}
G_{\bb{A}}(\bb{r},\bb{r}')\!=\!e^{\chi(\bb{r})\sigma_z +
i\varphi(\bb{r})}G_{0}(\bb{r},\bb{r'}) e^{\chi(\bb{r}')\sigma_z
-i\varphi(\bb{r}')},
\e
where $G_0$ refers to the zero-energy Green function associated with the
Hamiltonian $H_0= v \bb{\sigma p}$. The phases $\varphi(\bb{r})$ and
$\chi(\bb{r})$ 
satisfy
$\pa_x\varphi+\pa_y\chi=e A_x/c\hbar$ and
$\pa_y\varphi-\pa_x\chi=e A_y/c\hbar$.
In the Landau gauge, $\bb{A}\!=\!(0,B x\!+\!\phi c/2eL)$, and fixing 
$\varphi$ and $\chi$ by the requirement  $\chi(0)\!=\!\chi(L)\!=\!0$ (en\-suring that the boundary conditions at the graphene-lead interfaces are not
affected by the magnetic field), we find
\be
\chi(\bb{r}) =x(L-x)/2\ell_B^2, \quad \varphi(\bb{r}) = y\, (L/2\ell_B^2 + \phi/2L),
\e
where $\ell_B=(c\hbar/e B)^{1/2}$ is the magnetic length. 

Scalar impurities are described by the T-matrix $T=2\pi \ell_s$, where $\ell_s$
is a finite scattering length. Adatoms are characterized
by a T-matrix $T^c_\zeta =
\ell_a(1+\zeta\sigma_z\tau_z+\sigma_{-\!\zeta}\tau_-\exp{[}i\theta^c_\zeta{]}
+\sigma_\zeta\tau_+\exp{[}-i\theta^c_\zeta{]}$ that depends on the sublattice
index ($\zeta=1$ for the A and $\zeta=-1$ for the B sublattice) and a site ``color''
$c=-1,0,1$, encoding the Bloch phase at the impurity site. We use
$\sigma_\pm=(\sigma_x\pm i\sigma_y)/\sqrt{2}$ and $\tau_\pm=(\tau_x\pm
i\tau_y)/\sqrt{2}$ with $\sigma_{x,y,z}$ and $\tau_{x,y,z}$ being the Pauli
matrices in the sublattice and valley space, respectively. The phase
$\theta^c_\pm = \pm \alpha + 4\pi c/3$ depends in addition on the angle
$\alpha$ between the x-axis and the bond direction of the graphene lattice. 

In the wide-sample limit, $W\gg L$, we arrive at a general
expression \cite{EPAPS} for the conductance of the form
\be
\textrm{G}=\frac{4e^2}{h} \lt.\frac{\pa^2
\mathcal{F}}{\pa\phi^2}\rt|_{\phi=0}
=\frac{4e^2}{\pi h}\lt(W/L + \pi S \rt),
\label{Gfin}
\e
where $S$ can be interpreted as the sum of amplitudes corresponding to closed paths via impurity sites that go from the left lead to the right one and then back to the left lead (see Fig.~\ref{fig:zero-B}).  The amplitudes are given by products of free Green functions describing the propagation between impurities and T-matrices characterizing the scattering off each impurity. For scalar impurities, we find
\be
S = 4 \tr (  \hat{Y}_s\h M_+\hat{Y}_s M_-  -\hat{Y}^2 M_+ M_-),
\e
where $\hat{Y} =L^{-1}\diag (y_1,y_2,\dots ,y_N)$ is the diagonal matrix
consisting of $y$ components of the impurity coordinates,
$\hat{Y}_s=\hat{Y}+i\ell_s \sigma_y/2L$, and $M_\pm=(1\pm i\pi\ell_s
\hat{R}/2L)^{-1}$. The elements of the matrix $\hat{R}$ are given by 
\be
\label{R}
R_{nm} = e^{\chi(\bb{r}_n)\sigma_z}\!\!  \bpm \frac{1}{\sin(z_n+z_m^*)} &
\frac{1-\delta_{mn}}{\sin(z_n-z_m)} \\
\frac{1-\delta_{mn}}{\sin(z_n^*-z_m^*)} &  \frac{1}{\sin(z_n^*+z_m)} \epm
e^{\chi(\bb{r}_m)\sigma_z},
\e
where $z_n=\pi(x_n+iy_n)/2L$ and $\delta_{nm}$ is the Kronecker
symbol. Thus, the calculation of the conductance for a particular 
configuration of scalar impurities amounts to the inversion of a matrix of the
size $2N{\times} 2N$. In the case of adatoms, $S$ has a
similar structure \cite{EPAPS}.

The longitudinal conductivity is defined as $\sigma_{xx}= L
\textrm{G}/ W$,  where $W\gg L$. For the numerical analysis we fix $W=4 L$ and
plot $\sigma_{xx}$ as a function of the system size $L$, while keeping the magnetic
length $\ell_B$ and the average distance between
impurities $\ell_\textrm{imp}$ fixed. 

We begin the presentation of our results by briefly considering the regime of
zero magnetic field. Figure~\ref{fig:zero-B} displays the conductivity of
graphene with scalar impurities and adatoms for various impurity strengths. 
The case of scalar impurities is shown in Fig.~\ref{fig:zero-B}a. In the limit $\ell_s/\ell_\textrm{imp} \to \infty$, the
system belongs to the class DIII (with a Wess-Zumino term) and shows a
logarithmic scaling towards a ``supermetal'' (infinite-conductivity) fixed point
\cite{Ostrovsky10}. A finite value of $\ell_s$ breaks the chiral symmetry, thus
yielding the symmetry class AII (with a topological $\theta$-term). However, the
supermetallic behavior remains almost unchanged \cite{ostrovsky07}. 

A crossover due to symmetry breaking takes place also for the case of adatoms
in Fig.~\ref{fig:zero-B}b but the behavior of the conductivity is essentially different. 
As expected, Anderson localization sets in at long scales since the system belongs
to the conventional Wigner-Dyson symmetry class AI. 
For a fixed concentration of impurities, $\ell_\textrm{imp}^{-2}$, the localization
length is a non-monotonous function of the impurity strength parametrized by the
scattering length $\ell_a$. Indeed, in the limit $\ell_a\ll \ell_\textrm{imp}$,
the mean free path is large, so that the system remains ballistic up to large
distances. The opposite limit, $\ell_a/\ell_\textrm{imp} \to \infty$,
corresponds to the case of vacancies that preserve a chiral symmetry of the
Hamiltonian. The system in this case belongs to the chiral class  BDI and its
conductivity remains finite (no localization) in the limit $L\to \infty$
\cite{Ostrovsky10}.  
Thus, there exists an intermediate value of the ratio $\ell_a/\ell_{\rm imp}$
for which the localization is the strongest. For large (but finite) values of 
$\ell_a/\ell_{\rm imp}$, the system behaves as chiral up to a certain scale due
to the vicinity to a fixed point of class BDI, and then gets attracted by the
localization fixed point of class AI. We choose $\ell_a/\ell_{\rm imp}=50$ to
illustrate the behavior of the system in magnetic field. 

While our results showing that the localization exists for a
generic disorder but disappears in the chiral limit (zero
energy and $\ell_a/\ell_\textrm{imp} \to \infty$) are consistent with
NL$\sigma$M, they are at variance with numerical works
\cite{Yuan10,roche-vacancies,mayou}. Apparently, numerical approaches used in
these papers were not sufficient to reliably explore quantum interference
effects at the Dirac point.

\begin{figure}
\centerline{\includegraphics[width=\columnwidth]{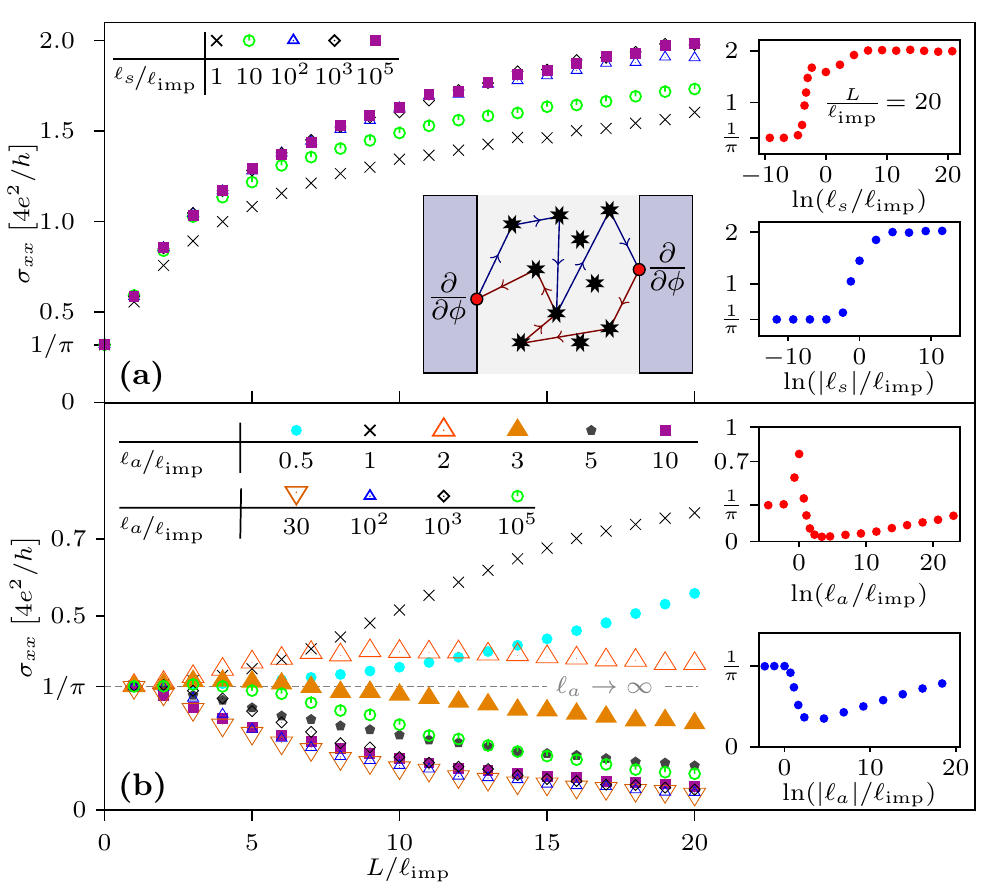}}
\caption{(Color online) Dirac-point conductivity of graphene with (a) scalar impurities 
and (b) adatoms in zero magnetic field as a function of the system size $L$. 
Insets show the dependence of $\sigma_{xx}$ on the impurity strength $\ell_a$ and
$\ell_s$ for a fixed system size $L=20 \ell_{\textrm{imp}}$. In the upper insets all impurities have the same sign while
in the lower insets the sign of the impurity potential is random. The sketch in (a) shows a typical path contributing to the conductance correction; this closed path connects the left with the right lead along impurity sites (\ding{88}) which are characterized by their individual T-matrices.
\vspace*{-0.3cm}}
 \label{fig:zero-B}
\end{figure}

\begin{figure*}
\centerline{\includegraphics[width=2\columnwidth]{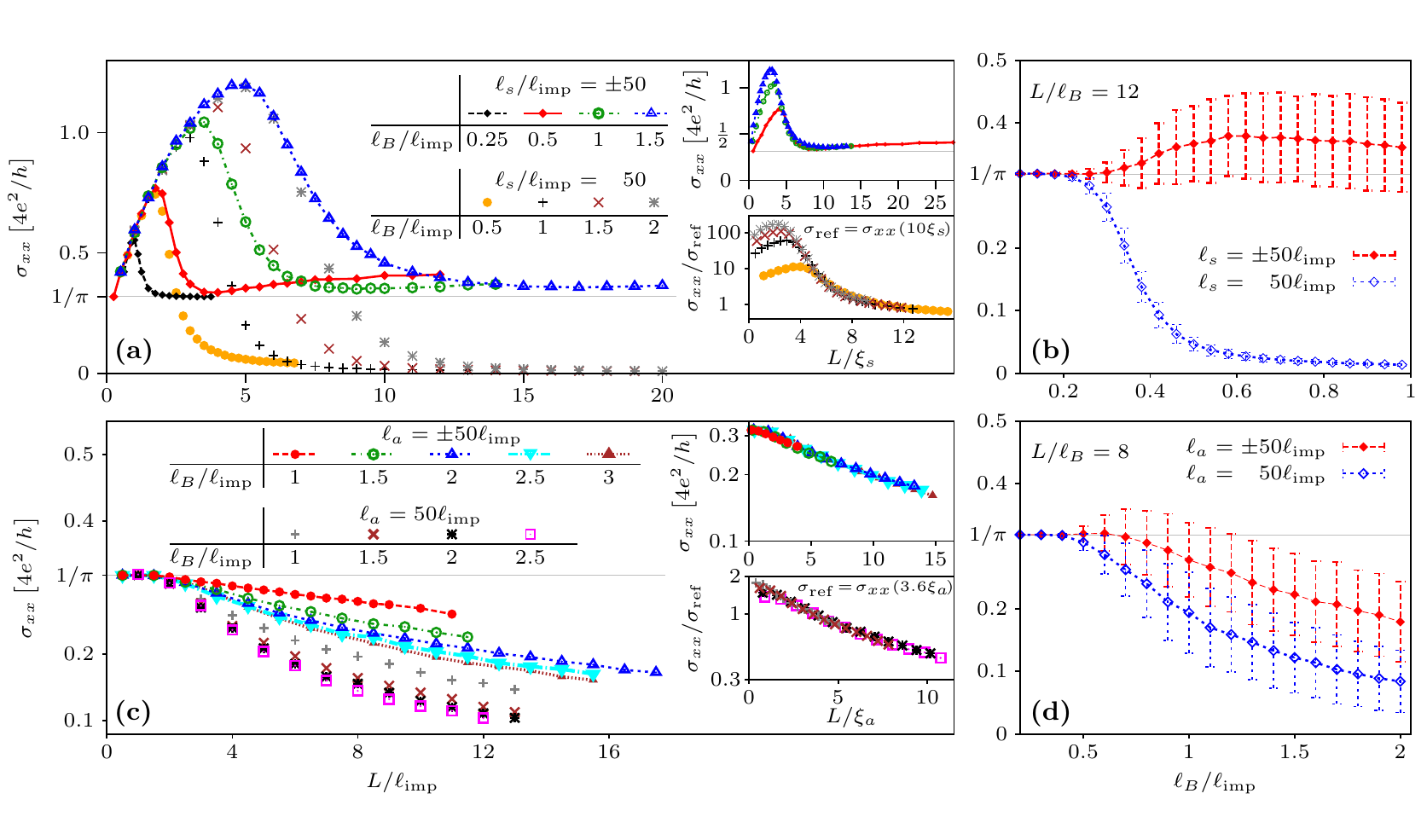}
}
\caption{(Color online) 
Zero-energy conductivity $\sigma_{xx}$ of graphene with scalar impurities (a,b)
and adatoms (c,d). Left panels: evolution of  $\sigma_{xx}$ with length $L$ for
different values of the ratio  $\ell_B/\ell_{\rm imp}$ of the magnetic length to
the distance between impurities. The symmetry breaking pattern is
DIII$\to$AII$\to$A for weaker $B$ and DIII$\to$AIII$\to$A for stronger $B$ in
panel (a) and BDI$\to$AI$\to$A for weaker $B$ and  BDI$\to$AIII$\to$A for
stronger $B$ in panel (c). Scaling flow towards the QH critical point in (a) and
localization in (a,c) is shown in the insets by rescaling of the curves to a length
$\xi_s$ (respectively, $\xi_a$). For the considered range of parameters,
the obtained values of $\xi_s$ and $\xi_a$ are well
approximated by phenomenological expressions $\xi_s =1.17 \ell_B (1 + 0.126
\ell_\textrm{imp}/\ell_B)$ and $\xi_a = \ell_\textrm{imp} + 
8.85\,\exp(-1.73 \ell_B/\ell_\textrm{imp})$. For impurities of the same sign an additional rescaling of $\sigma_{xx}$ 
reflects the two-parameter scaling in magnetic field for $\theta\neq 0,\pi$.
Right panels:  $\sigma_{xx}$ as a function of $\ell_B/\ell_{\rm imp}$
for fixed large $L/\ell_B$. For moderately strong $B$ the system is either at
QH criticality or gets localized. For stronger $B$,  a
quasi-ballistic transport regime with $\sigma = 4 e^2/\pi h$ emerges.
Vertical bars show mesoscopic fluctuations. 
\vspace*{-0.3cm}}
 \label{fig:main}
\end{figure*}

We are now in a position to turn to the case of strong magnetic field. Our main
results are shown in Fig.~\ref{fig:main}. For scalar impurities of random sign  
the center of the zeroth Landau level remains at the Dirac point, $E=0$, where the conductivity is calculated.
Different curves correspond to different strengths of the magnetic field $B$
parameterized by $\ell_B$. For small system sizes $L$, all curves follow the same supermetallic scaling (class AII)
characteristic for zero-$B$ case. When the magnetic field becomes important, the
symmetry class changes. While for infinitely strong impurities this would be the
chiral class AIII with the Wess-Zumino term, the finite value of $\ell_s$ places
the system into the class A with a $\theta=\pi$ topological term
\cite{Ostrovsky08}. This implies that the system should flow into the
QH critical point, see upper inset in Fig.~\ref{fig:main}a.  
The obtained value of the QH critical conductivity is $\sigma_* \simeq 4\times 0.4
e^2/h$, where the factor four is the total (spin and valley) degeneracy of the
scalar-impurity model. Remarkably, this critical value is approached from the
bottom.  This value is close to the one obtained recently in
Ref.~\cite{Ortmann13} for a tight-binding model of graphene with box disorder at
all sites.  Evidence of quantum Hall criticality of
Dirac fermions with long-range disorder was also reported in
Ref.~\cite{nomura08} where the Thouless number was numerically evaluated.

When all impurities are of the same sign, the critical state of the lowest
Landau level is shifted from the zero-energy point. In terms of the NL$\sigma$M
theory this implies that the topological term has now a prefactor $\theta$
different from $\pi$. The system should then scale towards $\sigma=0$ due to
Anderson localization. This is indeed seen in Fig.~\ref{fig:main}a (the main panel and lower inset). 
It is worth emphasizing that this localization in the QH regime is much more efficient that in zero $B$
(see Fig.~\ref{fig:zero-B}). 

Figures \ref{fig:main}b demonstrates another peculiarity of the problem. 
When the magnetic field is sufficiently strong such that the number of flux quanta
$N_\Phi$ exceeds $4N$ where $N$ is the number of impurities,  the conductivity
is given by its ballistic value $\sigma_{xx}=4e^2/\pi h$ independent of the
impurity positions. The condition $N_\Phi > 4N$, which translates into $\ell_B
< \ell_\textrm{imp}/2\sqrt{2\pi}$, can be understood from the following
argument. Each point-like scalar impurity cannot broaden the entire Landau level
but rather reduces its degeneracy by splitting four levels \cite{Avishai93}.
Indeed, the wave-functions of the degenerate Landau level can be superimposed such that
their values at the impurity sites are zero. Thus, for $N < N_\Phi/4$, a
macroscopic degeneracy of the Landau level remains, with the corresponding
eigenstates unaffected by  the  impurities. 
As a result the system does not flow to either QH or
localization fixed points but rather stays essentially ballistic. When  $N\ll
N_\Phi$, we obtain the ``ballistic conductivity'' $\sigma=4e^2/\pi h$ with
exponentially suppressed fluctuations. 

Figures \ref{fig:main}c and \ref{fig:main}d show the
behavior of the conductivity in the presence of adatoms. The limit of infinitely
strong adatoms (i.e., vacancies, $\ell_a\to\infty$) would correspond to the
chiral symmetry class AIII, which is characterized by a constant value of
$\sigma_{xx}$ close to $(4/\pi) e^2/h$. This is what we indeed observe at not
too large $L$ in Fig.~\ref{fig:main}c. For larger $L$ the chiral symmetry
breaking due to a finite $\ell_a$ occurs, and the system is in the symmetry
class A with a topological term. Contrary to the case of scalar impurities, we
observe localization (i.e., $\theta \ne \pi$) both for symmetric and asymmetric
distribution of random potential. This is because the intervalley scattering
splits the critical state of the lowest Landau level \cite{Ostrovsky08},
implying that the states at $E=0$ are now localized. 

Similarly to the model with scalar impurities, in a strong magnetic field we
observe the ballistic value of conductivity  $\sigma = (4/\pi) e^2/h$ with no
conductance fluctuations, as is seen in Fig.~\ref{fig:main}d. For adatoms, the
condition that a part of the Landau level eigenstates remain unaffected reads $N
< N_\Phi$, or equivalently, $\ell_B /\ell_\textrm{imp} < 1/\sqrt{2\pi}$. 

To summarize, we have studied the zero-energy conductivity of graphene with
strong (but not infinitely strong) short-range impurities in a magnetic field.
For this purpose, we have employed the unfolded scattering theory for the Dirac
Hamiltonian with point-like scatterers. The problem shows a complex scaling
behavior controlled by a number of fixed points reflecting the (approximate and
exact) symmetries as well as the topology of the problem. In the ultimate
long-length (low-temperature) limit, the system flows either into the QH
critical point or gets localized. The obtained value of the critical
conductivity is $\sigma_* \simeq 0.4 e^2/h$ times the degeneracy factor (equal
to four for scalar impurities). The localization at $E=0$ takes place when the
critical state of the zeroth Landau level is shifted by non-symmetric disorder, or
else, split by intervalley scattering on adatoms. When the magnetic field is so
strong that the number of flux quanta exceeds the number of impurities, the
conductivity  recovers its ballistic value $(4/\pi) e^2/h$. 

The work was supported by the EPSRC Doctoral Training Centre in Condensed Matter
Physics (S.G.) and the Scottish Universities Physics Alliance (W.R.H.), by the
Dutch Science Foundation NWO/FOM 13PR3118, by the EU network InterNoM, by DFG
SPP 1459, and by BMBF. M.T. is grateful to KIT for hospitality.

\newpage
\onecolumngrid
\setcounter{enumiv}{0} 
\setcounter{equation}{0} 

\vspace{1cm}
\centerline{\bfseries ONLINE SUPPORTING INFORMATION}

\begin{quote}
In the supporting information we provide the technical details which are missing in the main text of the Letter. In particular we introduce the unfolded scattering theory and derive the expression for the conductance as a function of impurity coordinates for the case of adatoms and scalar impurities in a magnetic field. 
\end{quote}

\maketitle

\maketitle

\section{Unfolded scatering theory}
\label{e_sec:unfolded}

\subsection{General formulation}

In this section we introduce the full-counting statistics and derive Eq.~(3) of the main text for the coductance as a function of impurity coordinates, which is used to obtain our main results.

Electronic properties of clean graphene are modeled by the Dirac Hamiltonian, $H_{\bb{A}}= v\, \bb{\sigma} (\bb{p} - e \bb{A}/c)$, where $\bb{\sigma}=(\sigma_x,\sigma_y)$, $\bb{p}$ is the two-dimensional momentum operator, $\bb{A}$ is the vector potential, $v\approx c/300\approx 10^6$m/s is the electron velocity and $c$ is the speed of light. The disorder potential is given by the superposition of individual impurity potentials,  $V_i(\bb{r})$, $i=1,\dots N$, with negligible overlap. Thus, the distance between impurities is assumed to be much larger than both the lattice constant and the decay length of the impurity potential.  

Te be more specific, we consider a rectangular graphene sample with periodic boundary conditions in $y$ direction ($0<y<W$) and open boundary conditions in $x$ direction ($0<x<L$). The latter correspond to highly doped graphene leads \cite{e_Tworzydlo06,e_Titov10}. At zero energy this geometry is equivalent to that of a Corbino disk as shown in Section \ref{e_sec:corbino} below. The conductance of the sample may be defined as the second derivative of a generating function with respect to the counting field \cite{e_Titov10,e_Ostrovsky10}
\be
\label{e_G}
\textrm{G} = \frac{4e^2}{h} \lt.\frac{\pa^2 \mathcal{F}}{\pa\phi^2}\rt|_{\phi=0}.
\e
The generating function, $\mathcal{F}(\phi)$, is symbolically represented as an operator trace
\be
\label{e_full}
\mathcal{F}(\phi) = \tr \ln G^{-1}, 
\e
where $G(\phi)$ is the Green's function operator which depends on the counting field $\phi$. Introduction of such a counting field normally requires an extension of the Green's function to the Keldysh (retarded-advanced) space. Different ways to define the counting fields for the present problem are discussed in Section~\ref{e_sec:counting} in great detail. 

It is convenient to decompose the generating function into the sum of two terms
\be
\label{e_FCSdeco}
\mathcal{F}(\phi) = \mathcal{F}_0(\phi) + \delta\mathcal{F}(\phi), 
\e
where $\mathcal{F}_0$ describes the full-counting statistics of a clean sample, $\mathcal{F}_0(\phi)=W \phi^2/2\pi L$ for $W\gg L$, while $\delta\mathcal{F}(\phi)$ represents the impurity contribution. 

Using the Dyson equation, one can formally relate the generating function to the bare Green's function $G_0(\phi)$ of a clean graphene sample with leads. This gives
\be
\mathcal{F}_0=\tr\ln G_0^{-1}, \qquad \delta\mathcal{F} =\tr\ln(1-V G_0),
\e
where the trace is still understood in the operator sense and the potential is given by the sum of individual impurity potentials, $V=\sum V_i$. With the help of the unfolded approach introduced in Ref.~\cite{e_Ostrovsky10} we can reduce the operator trace in the expression for $\delta\mathcal{F}$ to the matrix one which is taken in the unfolded (impurity) space. 

This procedure is most straightforward provided that the scattering on impurities is taken into account in the s-wave approximation. In this case we define $\hat{V}=\diag{(V_1,V_2,\dots , V_{N})}$ as a diagonal matrix in the impurity space, where $N$ is the total number of impurities in the sample. We also define the matrix operator $\hat{G}_0$ of the dimension $N\times N$ with identical entries, $(\hat{G}_0)_{ij}=G_0$, and use the operator identity
\be
\label{e_id}
\delta\mathcal{F} =\tr\ln(1-\hat{V} \hat{G}_0)=\tr\ln(1-\hat{T} (\hat{G}_0-\hat{g}))+\tr\ln(1-\hat{V} \hat{g}).
\e
Here we introduce the $T$-matrix operator 
\be
T=\hat{V}\frac{1}{1-\hat{g}\hat{V}},
\e
and the operator $\hat{g}=\diag{(g,g,\dots g)}$, where $g$ is the Green's function in an infinite graphene sample. The last term in Eq.~(\ref{e_id}) does not depend on the counting field and can be omitted in the calculation of transport properties. The operator trace in Eq.~(\ref{e_id}) can be reduced to the matrix trace in the impurity space by taking the limit of point-like impurities. This procedure is described in detail in Refs.~\cite{e_Titov10,e_Ostrovsky10}. As the result, one reduces the complex operator expression for $\delta\mathcal{F}$ to a matrix trace
\be
\label{e_FCS}
\delta\mathcal{F}(\phi) = \tr \ln \lt[ 1 - T \hat{G}_{\textrm{reg;}\phi}\rt],
\e
where $\hat{T}=(T_1, T_2, \dots, T_{N_\textrm{imp}})$ consists of individual impurity T-matrices and
$\hat{G}_{\textrm{reg;}\phi}$ is a matrix in the ``unfolded'' impurity space of the dimension $N$. The elements of $\hat{G}_{\textrm{reg;}\phi}$ are related to the Green's function $G_{\bb{A};\phi}$, associated with the Hamiltonian $H_{\bb{A}}$:  
\be
\label{e_reg}
(\hat{G}_{\textrm{reg}})_{nm} = \bc 
G_{\bb{A}}(\bb{r}_n,\bb{r}_m), \quad & n\neq m \\
\lim\limits_{\bb{r}\to\bb{r}_n} \lt[G_{\bb{A}}(\bb{r},\bb{r}_n) - g_{\bb{r}-\bb{r}_n}\rt],\; & n=m 
\ec.
\e
The Green's function retains a matrix structure in Keldysh, sublattice, and valley spaces.
The function $g_{\bb{r}}$ stands for the Green's function of an infinite system without disorder and at zero counting field. 

There are quite a few specific properties of the zero-energy state in clean graphene which originate in the chiral symmetry of the Hamiltonian, $\sigma_z H_{\bb{A}} \sigma_z = -H_{\bb{A}}$. One consequence of the symmetry is the relation between the retarded and advanced Green's functions at zero energy, $G^{R} = - \sigma_z  G^{A} \sigma_z$, in the absence of impurities.  Even though the chiral symmetry is generally violated by disorder, we may use this symmetry to conveniently define the counting field $\phi$ without resorting to the Keldysh space as
\be
\label{e_retarded}
(i0 - H_{\bb{A}}+\sigma_y\,\phi/2 L)G_{\bb{A};\phi}^R(\bb{r},\bb{r}') = \delta(\bb{r}-\bb{r}'),
\e
which corresponds to the shift of the vector potential, $A_y\to A_y + \phi\,c /2 e L$, inside the sample. The link between this definition and more common definitions of the counting field is explained in detail in Section~\ref{e_sec:counting}.

The Equation (\ref{e_retarded}) has to be supplemented with the open boundary conditions at the graphene-lead interfaces,
\be
\label{e_conditions}
\bpm 1 & 1\epm G_{\bb{A};\phi}^R(\bb{r}_0,\bb{r}') = \bpm 1 & -1\epm G_{\bb{A};\phi}^R(\bb{r}_L,\bb{r}') =0, 
\e
where $\bb{r}_0=(0,y)$ and $\bb{r}_L=(L,y)$. The unitarity constraint on the $T$-matrix ensures that the retarded and advanced components of the $T$-matrix are equal. Substituting the function $G_{\bb{A};\phi}^R$ in Eq.~(\ref{e_reg}) we construct the retarded matrix $\hat{G}^R_{\textrm{reg;}\phi}$. Using straightforward algebra, which is relegated to Section~\ref{e_sec:counting}, we rewrite Eq.~(\ref{e_FCS}) as 
\be
\label{e_deltaF}
\delta\mathcal{F}=\tr \ln \lt[1- \mathcal{L} \bpm T & 0\\ 0 & \bar{T}  \epm \mathcal{L}  \bpm \hat{G}^R_{\textrm{reg;} \phi} & 0\\ 0 & \hat{G}^R_{\textrm{reg;}-\phi}  \epm \rt], 
\e
where $\mathcal{L}=(\Sigma_z+\Sigma_y)/\sqrt{2}$ acts in the Keldysh space and $\bar{T}=-\sigma_z T\sigma_z$. 

For impurities preserving the chiral symmetry one finds $\bar{T}=T$, hence $\delta\mathcal{F} = K(\phi) + K(-\phi)$ with $K(\phi)=\tr \ln (1- T\hat{G}^R_{\textrm{reg;} \phi})$. For the scalar or mass impurities, $T=-\bar{T}$, one finds  $\delta\mathcal{F} = \tr \ln (1- T\hat{G}^R_{\textrm{reg;} \phi} T \hat{G}^R_{\textrm{reg;} -\phi})$.

Another specific feature of the zero-energy state is the existence of a non-unitary gauge transformation, which makes it possible to gauge away the entire magnetic field. This transformation can be formulated as
\be
\label{e_trans}
G_{\bb{A}}(\bb{r},\bb{r}') =  e^{\chi(\bb{r})\sigma_z + i\varphi(\bb{r})}G_0(\bb{r},\bb{r}') e^{\chi(\bb{r}')\sigma_z -i\varphi(\bb{r}')},
\e
where $G_0$ refers to the zero-energy Green's function associated with the Hamiltonian $H_0= v \bb{\sigma p}$. The phases $\varphi(\bb{r})$ and $\chi(\bb{r})$ possess even larger gauge freedom as the vector potential itself due to the differential relations: $\pa_x\varphi+\pa_y\chi=e A_x/c\hbar$ and $\pa_y\varphi-\pa_x\chi=e A_y/c\hbar$.

Let us now employ the Landau gauge in the form $\bb{A}=(0,B x+\phi c/2eL)$ and fix the phases $\varphi$ and $\chi$ from the requirement  $\chi(0)=\chi(L)=0$, which ensures that the boundary conditions (\ref{e_conditions}) at the graphene-lead interfaces are not affected by the magnetic field. As the result we find
\be
\chi(\bb{r}) =x(L-x)/2\ell_B^2, \quad \varphi(\bb{r}) = y\, (L/2\ell_B^2 + \phi/2 L),
\e
where $\ell_B^2=c\hbar/e B$ is the square of the magnetic length. Note, that the periodicity of $G_{\bb{A}}$ in the $y$-coordinate translates into the quasiperiodicity of $G_0$. A similar solution holds in Corbino geometry and the symmetric gauge as discussed in Section~\ref{e_sec:corbino}. 

The principal role of the free Green's function, $g_{\bb{r}}$, in Eq.~(\ref{e_reg}) is to provide a finite expression for diagonal terms of the matrix $\hat{G}_{\textrm{reg;}\phi}$. Here we take advantage of the zero-energy Green's function of an infinitely extended graphene, $g_{\bb{r}} = -i\, \bb{\sigma r}  / 2\pi r^2$, which is the same for both retarded and advanced components. The transformation (\ref{e_trans}) suggests that the Green's function $g_{\bb{r}}$ is not affected by the vector potential, hence the T-matrix has to be defined at zero field as well. (In fact the true Green's function of an infinite graphene sheet acquires a non-vanishing diagonal term in a finite magnetic field since the function $\chi(\bb{r})$ is always unbounded in an infinite sample. This complication is, however, not essential for our construction.)  

Solving Eqs.~(\ref{e_retarded},\ref{e_conditions}) in the limit $W\gg L$ we obtain
\be
\label{e_GRreg}
\hat{G}^R_{\textrm{reg;}\phi}= -\frac{i}{4L}e^{i \phi \hat{Y}/2} \hat{R}\, e^{-i \phi \hat{Y}/2} +\frac{\phi\,\sigma_y}{4\pi L}, 
\e
where $\hat{Y} = L^{-1}\diag (y_1,y_2,\dots ,y_N)$ is a diagonal matrix consisting of the $y$-components of the impurity coordinates. The elements of the matrix $\hat{R}$ are given by
\be
\label{e_R}
R_{nm} = e^{\chi(\bb{r}_n)\sigma_z}\!\!  \bpm \frac{1}{\sin(z_n+z_m^*)} & \frac{1-\delta_{mn}}{\sin(z_n-z_m)} \\
\frac{1-\delta_{mn}}{\sin(z_n^*-z_m^*)} &  \frac{1}{\sin(z_n^*+z_m)} \epm e^{\chi(\bb{r}_m)\sigma_z},
\e
where $z_n=\pi(x_n+iy_n)/2L$ and $\delta_{nm}$ stands for the Kronecker delta. In Equation (\ref{e_R}) we neglected unnecessary phases that are not entering the final result due to the gauge invariance. Note that, despite the multiplicative form of the transformation (\ref{e_trans}), the diagonal elements of $\hat{G}_{\textrm{reg;}\phi}$ (\ref{e_GRreg}) acquire an additive term $\phi\,\sigma_y/4\pi L$. Using the result (\ref{e_GRreg}) it is straightforward to differentiate with respect to $\phi$ to calculate the zero-energy conductance for any given impurity configuration and magnetic field. 

Below we consider impurities of two types: scalar impurities and ad-atoms. The former are described by the $T$-matrix $T=2\pi \ell_s$, where $\ell_s$ is a finite scattering length \cite{e_Titov10,e_Ostrovsky10}. The latter correspond to the T-matrix, $T^c_\zeta = \ell_a\lt(1+\zeta\sigma_z\tau_z+\sigma_{-\!\zeta}\tau_-e^{i\theta^c_\zeta}+\sigma_\zeta\tau_+e^{-i\theta^c_\zeta}\rt)$, that depends on the sublattice index ($\zeta=1$ for A and $\zeta=-1$ for B sublattice) and a site "color" $c=-1,0,1$, which encodes the Bloch phase at the impurity site. We use $\sigma_\pm=(\sigma_x\pm i\sigma_y)/\sqrt{2}$ and $\tau_\pm=(\tau_x\pm i\tau_y)/\sqrt{2}$ with $\sigma_{x,y,z}$ and $\tau_{x,y,z}$ being the Pauli matrices in the sublattice and valley space, respectively. The phase $\theta^c_\pm = \pm \alpha + 4\pi c/3$ depends in addition on the angle $\alpha$ between the $x$-axis and the bond direction of the graphene lattice. 
 
Using Eqs.~(\ref{e_G},\ref{e_full},\ref{e_retarded},\ref{e_GRreg}) we arrive at the general expression for the conductance in the form of Eq.~(3) of the main text
\be
\textrm{G}=\frac{4e^2}{\pi h}\lt(W/L + \pi S \rt).
\label{e_Gfin}
\e

In the case of scalar impurities we find
\be
\label{e_Sscalar}
S = 4 \tr (  \hat{Y}_s\h M_+\hat{Y}_s M_-  -\hat{Y}^2 M_+ M_-),
\e
where $M_\pm=(1\pm i\pi\ell_s \hat{R}/2L)^{-1}$  and $\hat{Y}_s=\hat{Y}+i\ell_s \sigma_y/2L$. Thus, the calculation of the conductance for a particular impurity configuration amounts to an inversion of a matrix of the size $2N\times 2N$.  

In the case of adatoms we find 
\be
\label{e_adS}
S =\tr \big\{ [\hat{Y},Q_+][\hat{Y},Q_-]
+ Q_+ [\hat{Y},\Gamma_+] Q_-[\hat{Y},\Gamma_-] + Q_+ [\hat{Y},\Gamma_-] Q_-[\hat{Y},\Gamma_+] \big\},
\e
where $[\;,\;]$ stands for the matrix commutator. We also introduced the matrices $\Gamma_\pm = (i\ell_a/8L) \hat{\zeta} A_\pm$, $Q_\pm = \lt[1\pm (\Gamma_++\Gamma_-) \rt]^{-1}$, where $\hat{\zeta} =\diag(\zeta_1,\zeta_2,\dots,\zeta_N)$ is a diagonal matrix in the impurity space consisting of sign factors, while the elements of the matrix $A_\pm$ are given by
\be
(A_\pm)_{nm}= \frac{e^{\pm\lt(\zeta_n \chi(x_n)+\zeta_m\chi(x_m)+i (\theta_n-\theta_m)/2\rt)}}
{\sin\frac{\pi}{2L} \lt[ \zeta_n x_n+\zeta_m x_m +i(y_n-y_m)\rt]},
\e
where the phases $\theta_n$ take different values, specified above, depending on the lattice orientation and the color of the corresponding atomic site. The calculation of the conductance using Eqs.~(\ref{e_Gfin},\ref{e_adS}) amounts to an inversion of matrix of the size $N\times N$ for each impurity realization. 

The detailed derivation of Eqs.~(\ref{e_Sscalar},\ref{e_adS}), which are used for numerical simulations of the conductivity, is given in the subsections \ref{e_subsec:scalar} and \ref{e_subsec:adatoms}.

\subsection{Numerical simulation of conductivity and its connection to quantum-Hall RG flow}

The two-dimensional longitudinal conductivity is defined as $\sigma_{xx}= L \textrm{G}/ W$ assuming the limit $W\gg L$. For numerical analysis we use $W=4 L$ and study the conductivity as a function of the system size $L$. The magnetic length $\ell_B =\sqrt{c\hbar/eB}$ and the average distance between impurities $\ell_\textrm{imp}$ remain fixed. We calculate the conductance for a number of randomly chosen impurity configurations, typically about 2000, and average to obtain the mean conductivity.
 
We focus on four distinct cases: i) scalar impurities represented by equivalent positive potentials corresponding to the impurity length scale $\ell_s=50 \ell_\textrm{imp}$; ii) scalar impurities represented by potentials of the same strength but a random sign, such that half of the impurities are modeled by positive potentials and another half by negative ones. We refer to this case symbolically as  $\ell_s=\pm 50 \ell_\mathrm{imp}$; iii) adatoms of random color and the same sign, $\ell_a=50 \ell_\mathrm{imp}$; and iv) adatoms of random color and random signs $\ell_a= \pm 50 \ell_\mathrm{imp}$.

In the presence of sufficiently strong magnetic field such that $\ell_\mathrm{imp} \ll \ell_B\ll L$ we may expect the conductivity to ``flow'' with the system size in accordance with the RG flow diagram depicted in Fig.~\ref{e_fig:RG} \cite{e_ostrovsky07}. The case of scalar impurities corresponds to the diagram shown in Fig.~\ref{e_fig:RG}a. When all impurities are modeled by positive potentials, the zeroth Landau level is broadened and shifted from zero energy. Therefore, zero energy, where we study transport, no longer corresponds to the half-filled Landau level. The RG flow in this case is schematically shown by the blue arrow at the left panel and localization behavior is expected. Scalar impurities modeled by potentials with alternating sign lead to a symmetric broadening of the zeroth Landau level. Zero energy, in this case, corresponds to the half-filled Landau level. The respective flow is indicated by the red arrow in Fig.~\ref{e_fig:RG}a. The longitudinal conductivity in this case is expected to flow to the quantum-Hall critical point $\sigma_\ast=2 g_U^*$, which is approximately given by $0.4\times 4e^2/h$ for graphene. The different types of behavior are indeed observed in the numerical simulation as shown in the top panels of Fig.~2 of the main text. 

Such a difference is absent at zero magnetic field. Antilocalization behavior is obtained irrespective of the sign of the scalar potentials. The behavior of conductance depicted in Fig.~\ref{e_fig:B0alternating}a for the case of alternating scalar impurities is qualitatively the same as that for the case of positive scalar impurities shown in the Letter.

\setcounter{figure}{2}

\begin{figure}
\centerline{\includegraphics[width=0.9\columnwidth]{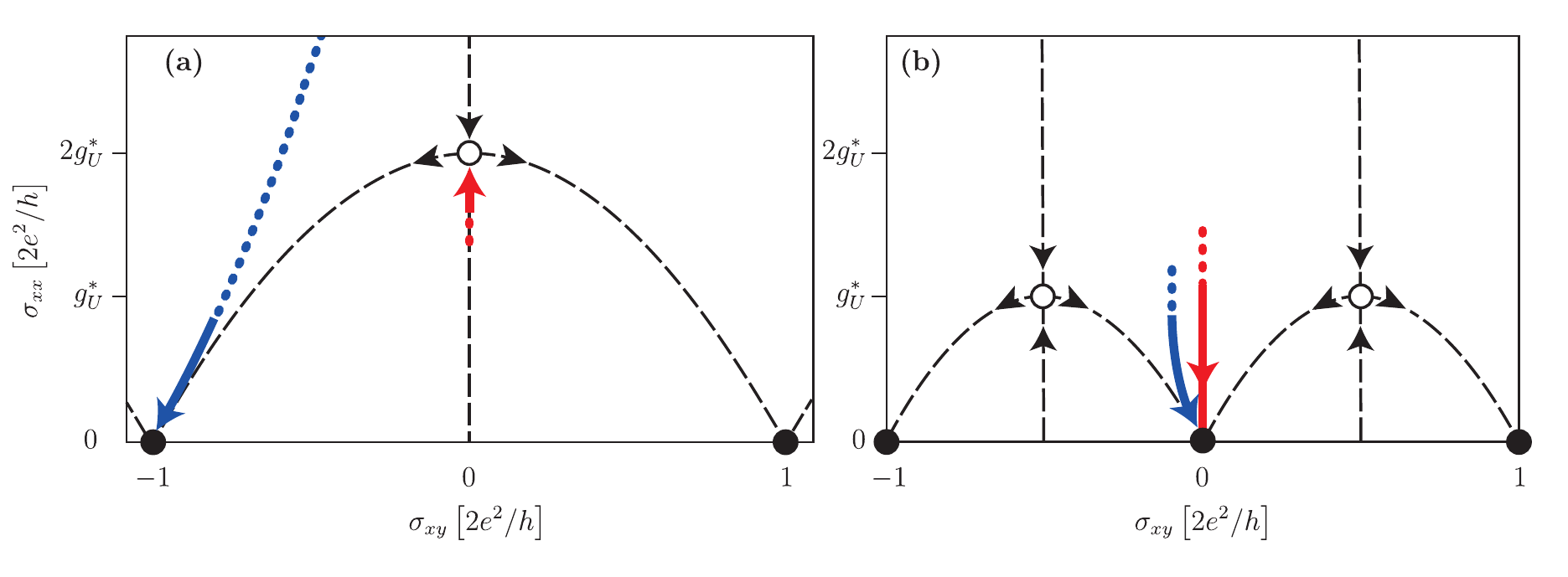}}
\caption{Schematic presentation of RG flow for (a) scalar impurities and (b) adatoms. Red lines illustrate the flow for the case of impurities of random sign while blue lines correspond to equal-sign impurities. 
\vspace*{-0.3cm}}
 \label{e_fig:RG}
\end{figure}

In the case of adatoms, localization is expected for both cases: at half-filling (for adatoms modeled by potentials of alternating signs) which corresponds to the red arrow in Fig.~\ref{e_fig:RG}b, and at a small detuning from the half-filling (for adatoms modeled by potentials of the same sign) which corresponds to the blue arrow in Fig.~\ref{e_fig:RG}b. This is confirmed by numerical simulations (see the lower panels of Fig.~2 of the main text).

The mean conductivity for the case of scalar impurities and adatoms of random sign is shown in Fig.~\ref{e_fig:B0alternating}b for completeness. 

\subsection{Scalar impurities}\label{e_subsec:scalar}

Let us consider scalar impurities or other impurities such that $T \sigma_z = \sigma_z T$. In this case, Eq.~(\ref{e_deltaF}) simplifies to 
\be
\label{e_scal1}
\delta\mathcal{F} = 2 \tr \ln (1-T \hat{G}^R_{\textrm{reg;}\phi} T \hat{G}^R_{\textrm{reg;}-\phi}),
\e
where  $\hat{G}^R_{\textrm{reg;}\phi}$ is given by Eq.~(\ref{e_GRreg}) and the factor of $2$ appears due to the trace over the valley degree of freedom. For scalar impurities we find $T= 2\pi \ell_s$, where $\ell_s=\diag(\ell_{s,1},\ell_{s,2},\dots \ell_{s,N})$ is regarded as a diagonal matrix in the impurity space to allow for different scattering lengths of different impurities. The matrix $T$ is proportional to the unit matrix in the sublattice space. 

In order to derive the expression for the conductance  we take advantage of a cyclic permutation under the trace to rewrite Eq.~(\ref{e_scal1}) as 
\be
\label{e_scal2}
\delta\mathcal{F} = 2 \tr \ln (1-\Gamma(\phi) \Gamma(-\phi)), \quad\mbox{where}
\qquad
\Gamma(\phi)=e^{i Y \phi} (-i\pi s R + \phi s \sigma_y), \qquad s\equiv \frac{\ell_s}{2L}.
\e
The matrix $R$ is defined in Eq.~(\ref{e_R}). We also use the fact that the following matrices commute, $[s,Y]=[s,\sigma_y]=[Y,\sigma_y]=0$.  Thus, the impurity correction to the conductance $\delta \textrm{G}$ reads
\be
\label{e_sim3}
\delta \textrm{G}= g_0 \lt.\frac{\pa^2 \delta \mathcal{F} }{\pa\phi^2}\rt|_{\phi=0} =4 g_0  \tr \lt.\lt(\frac{1}{1+\Gamma} \dot{\Gamma} \frac{1}{1-\Gamma}\dot{\Gamma}-\frac{\Gamma}{1-\Gamma^2}\ddot{\Gamma}\rt)\rt|_{\phi=0},\qquad g_0\equiv \frac{4e^2}{h},
\e
where we find from Eq.~(\ref{e_scal2})
\beml
\label{e_dots}
\beq
\Gamma &=& \Gamma(0) = - i \pi s R ,\\
\dot{\Gamma} &=& \lt.\frac{\pa \Gamma(\phi)}{\pa \phi}\rt|_{\phi=0}= i Y \Gamma+s\sigma_y,\\
\ddot{\Gamma} &=& \lt.\frac{\pa^2 \Gamma(\phi)}{\pa \phi^2}\rt|_{\phi=0} =2i Y s\sigma_y -Y^2 \Gamma. 
\eq
\eml
Substituting Eqs.~(\ref{e_dots}) in Eq.~(\ref{e_sim3}), one obtains
\be
\label{e_sim4}
\frac{\delta \textrm{G}}{g_0}=4  \tr \lt[ Y^2\frac{\Gamma^2}{1\!-\!\Gamma^2}- Y\frac{\Gamma}{1\!-\!\Gamma}Y \frac{1}{1\!+\!\Gamma} +\frac{1}{1\!+\!\Gamma} s\sigma_y \frac{1}{1\!-\!\Gamma} s\sigma_y+ i s \sigma_y \lt(\frac{1}{1\!-\!\Gamma}Y\frac{\Gamma}{1\!+\!\Gamma} + \frac{1}{1\!+\!\Gamma}Y\frac{\Gamma}{1\!-\!\Gamma} -2 Y \frac{\Gamma}{1\!-\!\Gamma^2}\rt) \rt].
\e
Using trivial identities like
\be
\frac{\Gamma}{1+\Gamma} = 1-\frac{1}{1+\Gamma}, \qquad \frac{\Gamma}{1-\Gamma} = -1+\frac{1}{1-\Gamma},
\e
we reduce Eq.~(\ref{e_sim4}) to
\be
\label{e_sim5}
\delta \textrm{G}= 4 g_0 \tr \lt((Y-is\sigma_y) \frac{1}{1-\Gamma}(Y+i s\sigma_y) \frac{1}{1+\Gamma} - Y^2\frac{1}{1-\Gamma^2}\rt),
\e
which is equivalent to Eqs.~(\ref{e_Gfin},\ref{e_Sscalar})

\subsection{Adatoms}\label{e_subsec:adatoms}

In the case of ad-atoms, the $T$-matrix is mixing valleys, hence we also have to care about the valley space. The $T$-matrices, $T_\pm$, for $A$ and $B$-type ad-atoms read
\be
T_+ = \frac{\ell_a}{2} 
\bpm 1 & 0& 0& e^{-i\theta_+^c}\\ 0& 0& 0 & 0\\ 0& 0& 0 & 0\\ e^{i\theta_+^c} & 0 & 0& 1\epm_{\sigma\tau}, 
\qquad T_-=\frac{\ell_a}{2}  
\bpm 0& 0& 0 & 0\\ 0& 1 &  e^{-i\theta_+^c}& 0 \\ 0& e^{i\theta_+^c} & 1 & 0 \\ 0& 0& 0 & 0\epm_{\sigma\tau}.
\e
These matrices are acting in the sublattice-valley space. The phase $\theta$ is determined by the ``color'' of the corresponding adatom site, $c=-1,0,1$, and the angle $\alpha$ between the carbon bond and the $x$-axis, $\theta^c_\pm= \pm \alpha +4\pi c/3$. In what follows we simply numerate the phases $\theta$ by the impurity index and construct the diagonal matrix $\theta=\diag(\theta_1,\theta_2,\dots, \theta_{N_\textrm{imp}})$ in the unfolded space. It is important to understand that the $T$-matrix is proportional to a projector in the sublattice-valley space. Using the rotation matrix,
\be
U=\frac{1}{\sqrt{2}} \bpm 1 & 0& 0& e^{-i\theta}\\ 0& 1 &  e^{-i\theta}& 0\\  0& e^{i\theta} & -1 & 0\\ e^{i\theta} & 0 & 0& -1\epm=U\h,
\e
we find
\be
U\h T_\pm U=\ell_a P_\pm, 
\e
where $P_\pm$  are simple projectors
\be
P_+=\bpm 1&0&0&0\\  0&0&0&0\\ 0&0&0&0\\ 0&0&0&0 \epm,\qquad
P_-=\bpm 0&0&0&0\\  0&1&0&0\\ 0&0&0&0\\ 0&0&0&0 \epm.
\e
The rotation $U$ can be applied to the Green's function in Eq.~(\ref{e_deltaF}) so that
\be
\delta\mathcal{F} = \tr \ln \lt(1-\ell_a P \lt[U G U\h\rt]  \rt),
\e
where we defined
\be
\label{e_old}
G_{nm}=\Lambda e^{i\Sigma_y\frac{\phi}{2L}(y_n-y_m)}G^{R}_{nm}\Lambda, \qquad G^{R}_{nm}=-\frac{i}{4L}R_{nm}++\frac{\phi\sigma_y}{4\pi L}\delta_{nm}.
\e
It is now necessary to calculate the elements of the matrix $U G U\h$ in the impurity space. This calculation gives
\beq
\n &&U_1\h G U_2
= \frac{1}{2} \bpm G+\sigma_x G\sigma_x e^{-i(\theta_1-\theta_2)} & G\sigma_x e^{-i\theta_2}-\sigma_x Ge^{-i\theta_1}\\
\sigma_x G e^{i\theta_1}-G\sigma_x e^{i\theta_2} & G+\sigma_x G\sigma_x e^{i(\theta_1-\theta_2)} \epm\\
&&= \frac{1}{2} \bpm e^{-i\theta_1/2} & 0\\ 0 & \sigma_x e^{i\theta_1/2}\epm 
\bpm  G e^{i\theta_{12}/2} +\sigma_x G\sigma_x e^{-i\theta_{12}/2} & G e^{i\theta_{12}/2} - \sigma_x G\sigma_x e^{-i\theta_{12}/2} \\
G e^{i\theta_{12}/2} - \sigma_x G\sigma_x e^{-i\theta_{12}/2}  &  Ge^{i\theta_{12}/2}+\sigma_xG \sigma_xe^{-i\theta_{12}/2}\epm \bpm e^{i\theta_2/2} & 0\\ 0 & \sigma_x e^{-i\theta_2/2}\epm,
\eq
where $\theta_{12}\equiv\theta_1-\theta_2$ and the matrix structure in the valley space is explicitly shown. Note that due to the projective properties of $P$ we only need to know the upper left element in the valley space!  It is, therefore, convenient to introduce the valley-reduced matrix $\bar{G}$  with the elements 
\be
\bar{G}_{nm} =\frac{1}{2}\lt(  G_{nm} e^{i(\theta_{n}-\theta_m)/2} +\sigma_x G_{nm} \sigma_x e^{-i(\theta_{n}-\theta_m)/2} \rt)
\e
which can be used to rewrite the result for the full-counting statistics as 
\be
\label{e_res2}
\delta\mathcal{F} = \tr \ln \lt[1-\ell_a P \bar{G} \rt],\qquad P_+=\bpm 1& 0\\ 0 & 0\epm, \qquad P_-=\bpm 0& 0\\ 0 & 1\epm.
\e
At the next step we are going to calculate the trace in the Keldysh space. First of all we note that $\sigma_x \Lambda\sigma_x = \Lambda\Sigma_z$, hence
\be
\sigma_x G_{nm} \sigma_x = \Lambda \Sigma_z e^{i\Sigma_y(y_n-y_m)\phi/2L} \Sigma_z G^R_{nm} \Lambda = \Lambda e^{-i\Sigma_y(y_n-y_m)\phi/2L} G^R_{nm} \Lambda.
\e
Using that $\Lambda P \Lambda = \Sigma_z P$ we obtain
\beq
\delta\mathcal{F} &=& \tr \ln \lt[1- \ell_a P \Sigma_z \frac{1}{2} \lt(e^{i\Sigma_y \frac{\phi(y-y')}{2L}+i\frac{\theta-\theta'}{2}}G^R+e^{-i\Sigma_y \frac{\phi(y-y')}{2L}-i\frac{\theta-\theta'}{2}}\sigma_xG^R\sigma_x\rt)\rt]\\
&=&\tr \ln \lt[1- \ell_a P \Sigma_y \frac{1}{2}\lt(e^{i\Sigma_z \frac{\phi(y-y')}{2L}+i\frac{\theta-\theta'}{2}}G^R+e^{-i\Sigma_z \frac{\phi(y-y')}{2L}-i\frac{\theta-\theta'}{2}}\sigma_xG^R\sigma_x\rt)\rt]\\
&=&\tr\ln \lt[1-\ell_a^2 PS(\phi)PS(-\phi) \rt]
\eq
where we used self-explanatory symbolic notations in the parenthesis to make the expression more compact.  We have also defined
\be
S_{nm}(\phi) = \frac{1}{2}\lt(
e^{i\frac{\phi}{2L}(y_n-y_m)+\frac{i}{2}(\theta_n-\theta_m)} G_{nm}^R+
e^{-i\frac{\phi}{2L}(y_n-y_m)-\frac{i}{2}(\theta_n-\theta_m)} \sigma_xG_{nm}^R \sigma_x \rt).
\e
The projectors $P_\pm$ are just selecting appropriate elements of the matrix $S$. The final step is just to introduce more compact notations in order to eliminate the projectors. First of all we notice that the second term in the expression for $G^{R}$ in Eq.~(\ref{e_old}) never contributes. Indeed only diagonal elements of $S_{nn}(\phi)$ in the sub-lattice space enter the result. This is clearly what we expect in the case of ad-atoms on very general grounds. Thus we can simply use 
\be
G^R_{nm} = -\frac{i}{4L}
\bpm \frac{e^{\chi(x_n)+\chi(x_m)}}{\sin\lt[\frac{\pi}{2L}(x_n+x_m+i(y_n-y_m))\rt]} & 
\frac{e^{\chi(x_n)-\chi(x_m)}}{\sin\lt[\frac{\pi}{2L}(x_n-x_m+i(y_n-y_m))\rt]} \\ 
-\frac{e^{-\chi(x_n)+\chi(x_m)}}{\sin\lt[\frac{\pi}{2L}(-x_n+x_m+i(y_n-y_m))\rt]}& 
-\frac{e^{-\chi(x_n)-\chi(x_m)}}{\sin\lt[\frac{\pi}{2L}(-x_n-x_m+i(y_n-y_m))\rt]}\epm.
\e
Taking advantage of the sign convention ($\zeta=1$ for $A$-site and $\zeta=-1$ for B-site) we can write 
\beq
P_nG^R_{nm}P_m &=&-\frac{i}{4L} \frac{\zeta_ne^{\zeta_n\chi(x_n)+\zeta_m\chi(x_m)}}{\sin\lt[\frac{\pi}{2L}\lt(\zeta_nx_n+\zeta_mx_m+i(y_n-y_m)\rt)\rt]},\\
P_n\sigma_x G^R_{nm}\sigma_x P_m &=&-\frac{i}{4L} \frac{\zeta_ne^{-\zeta_n\chi(x_n)-\zeta_m\chi(x_m)}}{\sin\lt[\frac{\pi}{2L}\lt(\zeta_nx_n+\zeta_mx_m-i(y_n-y_m)\rt)\rt]}.
\eq
Thus, we obtain the final expression for the full counting statistics in the form
\be
\label{e_fullres}
\delta\mathcal{F}=\tr\ln\lt(1+\frac{\ell_a^2}{16L^2} \hat{\zeta} K(\phi) \hat{\zeta} K(-\phi)\rt),\qquad K(\phi)=\frac{1}{2}(A_+(\phi)+A_-(\phi)),
\e
where $\hat{\zeta}=\diag(\zeta_1,\zeta_2,\dots \zeta_N)$ is a diagonal matrix in the unfolded space and the matrices $A_+$ and $A_-$ are defined in the same way as in our previous publications
\be
\lt(A_{\pm}(\phi)\rt)_{nm}= \frac{e^{\pm\lt(\frac{i\phi}{2L}(y_n-y_m)+\frac{i}{2}(\theta_n-\theta_m)+\zeta_n\chi(x_n)+\zeta_m\chi(x_m)\rt)}}
{\sin\lt[\frac{\pi}{2L}(\zeta_nx_n+\zeta_m x_m\pm i(y_n-y_m))\rt]}.
\e 
In the vacancy limit, $\ell_a\to \infty$, we restore the known result $\delta\mathcal{F} = \tr \ln K(\phi) +\tr \ln K(-\phi)$ \cite{e_Ostrovsky10}. It is worth mentioning  that both $A_+$ and $A_-$ are Hermitian matrices and $A_-(B)=A_+^T(-B)$, where $B$ is the magnetic field. For a rectangular sample with $x\in(0,L)$ we use $\chi(x)=B x(L-x)/2$. 

The expression (\ref{e_fullres}) can be rewritten as
\be
\delta\mathcal{F}=\tr\ln\lt(1-\Gamma(\phi) \Gamma(-\phi)\rt),\qquad \Gamma(\phi)= -\frac{i\ell_a}{4L}\hat{\zeta} K(\phi).
\e
In order to find the conductance we  calculate the derivatives of $\Gamma$ at $\phi=0$
\be
\dot{\Gamma}=\frac{i}{2}[Y,\bar{\Gamma}],\qquad \ddot{\Gamma}=-\frac{1}{4}[Y,[Y,\Gamma]],
\e
where
\be
\Gamma=-\frac{i\ell_a}{4L}\hat{\zeta}\frac{A_++A_-}{2},\qquad \bar{\Gamma}=-\frac{i\ell_a}{4L}\hat{\zeta}\frac{A_+-A_-}{2}.
\e
The matrices $A_\pm=A_\pm(0)$ are now taken at $\phi=0$. Using Eq.~(\ref{e_sim3}) we obtain
\be
\lt.\frac{\pa^2 \delta \mathcal{F}}{\pa \phi^2}\rt|_{\phi=0} = 
\frac{1}{2}\tr 
\lt(\frac{\Gamma}{1-\Gamma^2} [Y,[Y,\Gamma]] - \frac{1}{1+\Gamma} [Y,\bar{\Gamma}] \frac{1}{1-\Gamma} [Y,\bar{\Gamma}] \rt),
\e
which is basically the final expression for the conductance. Using the identity
\be
\label{e_identity}
\tr  \lt(\frac{\Gamma}{1-\Gamma^2} [Y,[Y,\Gamma]] - \frac{1}{1+\Gamma} [Y,\Gamma] \frac{1}{1-\Gamma} [Y,\Gamma] \rt)=
2\tr [Y,\frac{1}{1+\Gamma}] [Y,\frac{1}{1-\Gamma}].
\e
we can also write
\be
\label{e_fin10}
\lt.\frac{\pa^2 \delta \mathcal{F}}{\pa \phi^2}\rt|_{\phi=0} =\tr \lt([Y,\frac{1}{1+\Gamma}] [Y,\frac{1}{1-\Gamma}] +\frac{1}{1+\Gamma}[Y,\Gamma_+]\frac{1}{1-\Gamma}[Y,\Gamma_-]+
\frac{1}{1+\Gamma}[Y,\Gamma_-]\frac{1}{1-\Gamma}[Y,\Gamma_+]\rt),
\e
where 
\be
\label{e_fin10b}
\Gamma_\pm=\frac{1}{2}(\Gamma\pm \bar{\Gamma})=-\frac{i\ell_a}{8L}\hat{\zeta} A_\pm. 
\e
The Equations~(\ref{e_fin10},\ref{e_fin10b}) are equivalent to Eqs.~(\ref{e_Gfin},\ref{e_adS}) given above.
In the vacancy limit, $\ell_a\to \infty$, we can simply disregard the first term in Eq.~(\ref{e_fin10}) and omit the unit matrix in the expressions $1\pm \Gamma$, hence
\be
\lim_{\ell_a\to\infty} \lt.\frac{\pa^2 \delta \mathcal{F}}{\pa \phi^2}\rt|_{\phi=0}  =  -  2\tr \frac{1}{\Gamma}[Y,\Gamma_+]\frac{1}{\Gamma}[Y,\Gamma_-] = - 2\tr \frac{1}{A_++A_-}[Y,A_+]\frac{1}{A_++A_-}[Y,A_-].
\e
This result is equivalent to the one obtained in Ref.~\onlinecite{e_Ostrovsky10} for the Dirac-point conductance of a rectangular graphene sample with $N$ vacancies.

\begin{figure}
\centerline{\includegraphics[width=0.9\columnwidth]{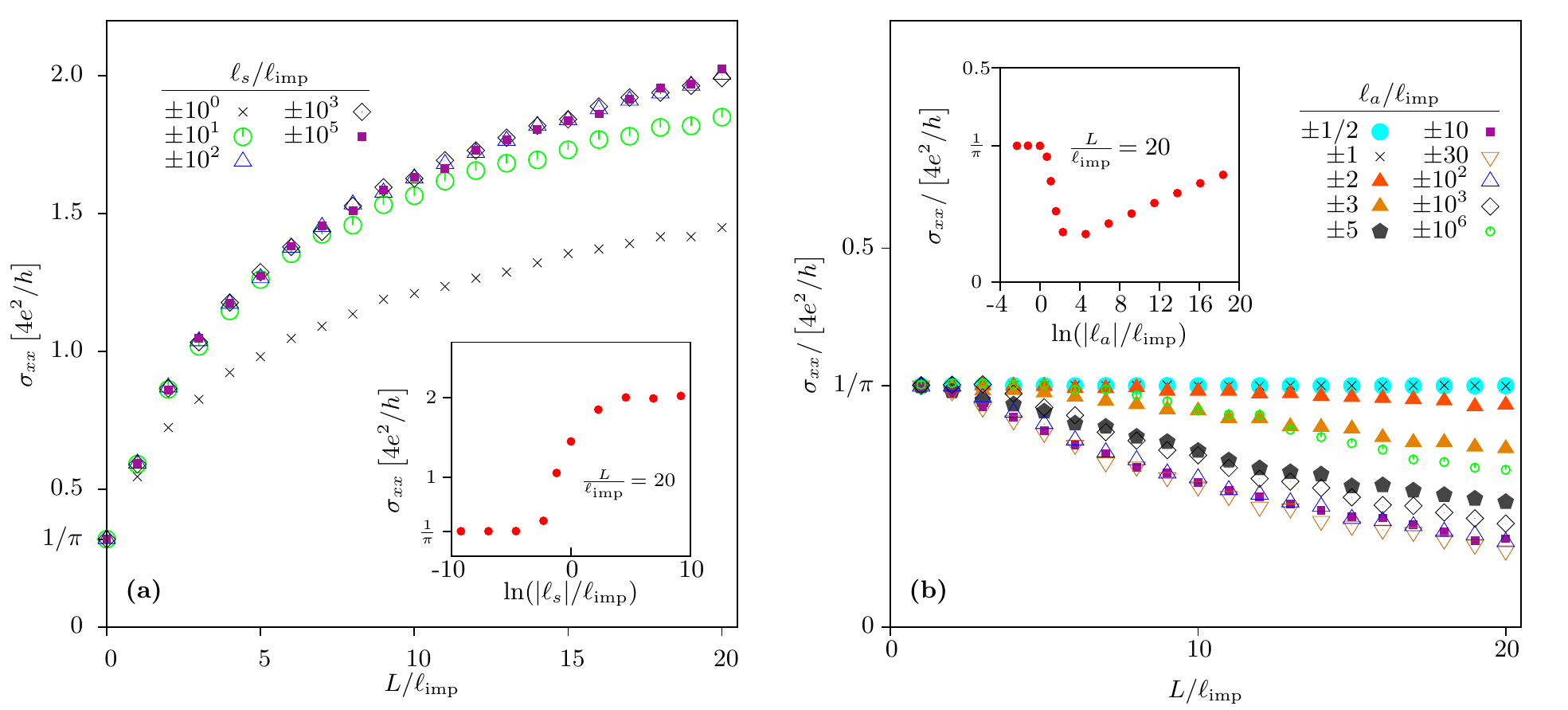}}
\caption{Conductivity in zero magnetic field as a function of system length for (a) scalar impurities and (b) adatoms of random sign.  The long-scale behavior is analogous to that for impurities of equal signs (see Fig. 1 of the main text). Insets show the conductivity as a function of impurity strength for fixed length and impurity concentration.
\vspace*{-0.3cm}}
 \label{e_fig:B0alternating}
\end{figure}

\section{Transformation of the counting fields}\label{e_sec:counting}

Transport quantities such as conductance, noise, and higher moments of charge transfer statistics can be directly inferred from the Green's function which depends on the counting fields. Following the general approach, we define the Keldysh Green's function for graphene as the solution of the following problem  
\be
\label{e_Kproblem3}
\bpm \ep+i0-H-\mu(x) & -\hbar v \sigma_x \zeta_+\,\delta(x) \\ -\hbar v \sigma_x\zeta_-\,\delta(x-L) & \ep-i0 -H -\mu(x) \epm G(\bb{r};\bb{r}')=\delta(\bb{r}-\bb{r}'),
\e
where we take $H=H_{\bb{A}}+ V(\bb{r})$, with $V(\bb{r})$ describing the impurity potentials and $H_{\bb{A}}=v \bb{\sigma}\lt(\bb{p}-e \bb{A}/c\rt)$. For the sake of definiteness, we impose periodic boundary conditions in $y$ direction, $y\in (0,W)$, and adopt open boundary conditions in $x$ direction, which correspond to infinitely-doped leads, i.e. $\mu(x<0)=\mu(x>L)= - \infty$ and $\mu(0<x<L)=0$. The counting fields $\zeta_\pm$ are introduced at the interfaces between the leads and the sample. This definition is largely conventional since any sample interface can be equivalently used to measure current.  

The Green's function $G(\bb{r};\bb{r}')$ contains information on the transport properties, which is encoded in the generating function  
\be
\label{e_def}
\mathcal{F}=\tr\ln G^{-1},
\e
where the trace includes the operator trace i.e. the integration over the spatial degrees of freedom. The conductance is, then, defined as the second derivative of the generating function with respect to the counting fields,
\be
\label{e_cond1}
\textrm{G} = - \frac{2e^2}{h}\lt.\frac{\pa^2 \mathcal{F}}{\pa\zeta_+\,\pa\zeta_-}\rt|_{\zeta_\pm=0},
\e
where the coefficient $2$ takes into account the spin degeneracy. (We generally assume the valley degeneracy to be lifted, hence the Green's function is regarded as a matrix in the valley space as well.)

Expression (\ref{e_cond1}) corresponds to the following definition of the conductance,
\be
\label{e_cond1a}
\textrm{G} = \frac{2e^2}{h} (\hbar v)^2 \int_0^W \!\!\!dy \int_0^W \!\!\!dy' \tr \sigma_x G^R(0,y; L,y') \sigma_x G^A(L,y';0,y), 
\e
where the Green's functions $G^{R,A}$ are found in the absence of the counting fields.

We can conveniently find the Green's function inside the sample, i.e. for $0<x<L$ or $a<r<R$, by solving the simpler equation
\be
\label{e_keld}
\bpm \ep+i0-H& 0 \\ 0 & \ep-i0 -H\epm G(\bb{r},\bb{r}')=\delta(\bb{r}-\bb{r'}),
\e
which is supplemented by the boundary conditions at the interfaces
\be
\label{e_bound}
\bpm 1 &1&i\zeta_+& i\zeta_+ \\ 0&0&1&-1\epm G(0,y;\bb{r}') =0,\qquad 
\bpm 1 &-1&0&0 \\ -i\zeta_-& -i\zeta_- &1&1\epm G(\bb{r};L,y') =0.
\e
Without loss of generality, the counting fields can be considered as equal $\zeta=\zeta_\pm$. To simplify the boundary conditions (\ref{e_bound}) for the Green's function, it is convenient to transform the counting field $\zeta$ to the field $\phi$ as
\be
\label{e_transform}
\zeta=i\sinh \phi/2,\qquad \textrm{G}
=-\frac{e^2}{h} \lt.\frac{\pa^2\mathcal{F}}{\pa\zeta^2}\rt|_{\zeta=0} 
=\frac{4e^2}{h}\lt.\frac{\pa^2\mathcal{F}}{\pa\phi^2}\rt|_{\phi=0}, 
\e
and adopt the unitary transformation of the Green's function in the Keldysh space
\be
\label{e_ups1}
G(\mathbf{r},\mathbf{r}')=V_\phi(x)  \Lambda_K \tilde{G}(\mathbf{r},\mathbf{r}')  \Lambda_K^{-1}  V_\phi^{-1}(x'),
\e
where
\be
V_{\phi}(x) = \frac{1}{\sqrt{2\cosh \phi/2}} \bpm e^{\frac{\phi(L-x)}{2L}} & -e^{-\frac{\phi(L-x)}{2L}} \\
e^{- \frac{\phi x}{2L}} & e^{\frac{\phi x}{2L}} \epm, \qquad
\qquad  \Lambda_K =\frac{1}{\sqrt{2}} \bpm 1 & 1 \\ -1 & 1\epm.
\e  
The transformation (\ref{e_ups1}) converts the problem (\ref{e_keld},\ref{e_ups1}) to an equivalent one, where the counting field is introduced in a slightly different manner,
\be
\label{e_keldysh3}
\bpm
\ep+i0-H & -i\hbar v\,\sigma_x \frac{\phi}{2L} \\
-i\hbar v\,\sigma_x \frac{\phi}{2L}  & \ep-i0- H
\epm \tilde{G}(\bb{r},\bb{r}')=\delta(\bb{r}-\bb{r}'),
\e
while the boundary conditions are simplified to
\begin{equation}
 \label{e_boundary2}
 \begin{pmatrix}
   1 & 1 & 0 & 0 \\
   0 & 0 & 1 & -1
 \end{pmatrix} \tilde{G}(0,y,\bb{r}')
  = 0,
 \qquad
 \begin{pmatrix}
   1 & -1 & 0 & 0 \\
   0 & 0 & 1 & 1
 \end{pmatrix} \tilde{G}(L,y;\bb{r}')
  = 0.
\end{equation}
The transformation (\ref{e_ups1}) holds at any energy and can be generalized to any Hamiltonian $H$ with arbitrary matrix potential $V$. Consequently the matrices $V_\phi$ and $\Lambda_K$ drop out completely in any physical observable.  The reason behind such a generality is merely the current conservation. Indeed, using the transformation Eq.~(\ref{e_transform}) one finds that the expression for conductance (\ref{e_cond1},\ref{e_cond1a}) can be rewritten as
\be
\textrm{G} =\frac{4e^2}{h}\lt.\frac{\pa^2 \mathcal{F}}{\pa\phi^2}\rt|_{\phi=0} =\frac{2e^2}{h} \frac{(\hbar v)^2}{L^2} \int \! d^2\bb{r} \!\int \! d^2\bb{r}' \,\tr \sigma_x G^R(\bb{r}; \bb{r}') \sigma_x G^A(\bb{r}';\bb{r}),
\e
where the integration extends over the sample area. This formula can be obtained directly from Eq.~(\ref{e_cond1a}) by taking advantage of the current conservation through an interface, which implies that the integrals of the type $\int dy\, G^R(\bb{r}_1;\bb{r}) \sigma_x G^A(\bb{r};\bb{r}_2) $, where $\bb{r}=(x,y)$, do not depend on $x$. 

Let us now consider the zero-energy Green's function for the Dirac hamiltonian in the absence of an impurity potential, $H=H_0$, so that the hamiltonian yields the chiral symmetry $H_0 \sigma_z+\sigma_z H_0 =0$. The symmetry suggests the relation between the retarded and advanced Green's functions at zero energy, $G_0^A=-\sigma_z G_0^R \sigma_z$, which makes the Keldysh space somewhat obsolete. The fermions at zero energy can be viewed as the their own antiparticles, which can be completely characterized by the retarded Green's function only. In this case we can further simplify the problem of Eqs.~(\ref{e_keldysh3},\ref{e_boundary2}) by using another transformation
\be
\label{e_ups2}
\tilde{G}_0(\mathbf{r},\mathbf{r}')=\Lambda \mathcal{L} \check{G}_0(\mathbf{r},\mathbf{r}')  \mathcal{L}^{-1} \Lambda,
\e
where the matrix $\mathcal{L}$ is a rotation in Keldysh space,
\be
\mathcal{L}= \frac{1}{\sqrt{2}}\lt(\Sigma_z+\Sigma_y\rt),\qquad \Lambda=\bpm 1 & 0 \\ 0 & i\sigma_z\epm.
\e
This transformation leads to the equation 
\be
\label{e_keldysh4}
\bpm
i0-H_0 +  \hbar v\,\frac{\phi}{2L}\sigma_y& 0 \\
0  & i0- H_0 - \hbar v\, \frac{\phi}{2L}\sigma_y
\epm \check{G}_0(\bb{r},\bb{r}')=\delta(\bb{r}-\bb{r}'),
\e
where we used the fact that the operator $H_0$ anticommutes with the matrix $\sigma_z$. It is easy to check that the solution to this equation must be diagonal in the Keldysh space, i.e.
\be
\check{G}_0(\bb{r},\bb{r}') = \bpm \check{G}^R_{0,\phi}(\bb{r},\bb{r}') & 0 \\ 0 & \check{G}^R_{0,-\phi}(\bb{r},\bb{r}') \epm,
\e
with the boundary conditions
\begin{equation}
 \label{e_boundary3}
 \begin{pmatrix}
   1 & 1 
 \end{pmatrix} \check{G}^{R}_{0,\pm \phi}(0,y,\bb{r}')
  = 0,
 \qquad
 \begin{pmatrix}
   1 & -1 
 \end{pmatrix} \check{G}^R_{0,\pm \phi}(L,y;\bb{r}')
  = 0.
\end{equation}
Thus, the equation (\ref{e_Kproblem2})  (or Eq.~(\ref{e_Kproblem3}) for the bare Green's function) is reduced to the solution of the retarded equation 
\be
\lt(i0-H_0 +\hbar v\, \frac{\phi}{2L}\sigma_y\rt) \check{G}^R_{0,\phi} (\bb{r},\bb{r}') = \delta(\bb{r}-\bb{r}'),
\e
where the counting field $\phi$ is simply added to the $y$-component of the vector potential. This equation holds for any boundary conditions in $y$-direction and any aspect ratio $W/L$. This approach can be generalized to arbitrary geometry using conformal mapping.

\section{Mapping between Corbino-disk and cylinder geometry}\label{e_sec:corbino}

\subsection{Corbino disk}

In this section we consider the mapping between the Corbino-disk geometry and the cylinder geometry for graphene.  We use the mapping, which can be regarded as a conformal mapping, to calculate the Dirac-point conductance of ballistic graphene in Corbino geometry subject to a perpendiclar magnetic field. 

Let us consider a graphene sample that has the shape of a Corbino disk with inner radius $a$ and outer radius $R$. The Dirac Hamiltonian inside the disk takes the form,
\be
H_\textbf{A} = v \bb{\sigma} (\bb{p} - e \bb{A}/c),
\e
where we use the symmetric gauge $\bb{A} = (-y, x) B/2$ for the sake of definiteness. It is convenient to take advantage of the polar coordinates $\bb{r}=(x,y)= (r\cos\theta,r\sin \theta)$ and place the origin to the center of the Corbino disk. Then, the following transformation applies
\be
\label{e_trans0}
e^{-(\ln r - i \sigma_z \theta)/2}H_\textbf{A}e^{-(\ln r + i \sigma_z \theta)/2} = r^{-2} \bar{H}_\textbf{A},
\e
where $\bar{H}_\textbf{A}=  v \bb{\sigma} (\bar{\bb{p}} - e \bar{\bb{A}}/c)$ has the form of the graphene Hamiltonian with
\be
\bar{p}_x = -i\hbar \,r \frac{\pa}{\pa r}, \quad \bar{p}_y = -i\hbar \frac{\pa}{\pa \theta}, \qquad \bar{\bb{A}} =(0, Br^2/2).
\e
It is convenient to introduce new dimensionless coordinates $\bar{x}=\ln r/a \in (0,L)$ and $\bar{y} =\theta \in (0,W)$, where $L=\ln R/a$ and $W=2\pi$. In these coordinates $\bar{H}_\textbf{A}$ describes a graphene cylinder with the length $L$ and the circumference $W$. Note, however, that the magnetic field is changing exponentially along the cylinder. In addition, the cylinder is pierced by the magnetic flux which equals to a half of the flux quanta $\Phi_0/2$, where $\Phi_0=hc/e$. This flux originates in the term $ \exp( i \sigma_z \theta /2)$ in the transformation (\ref{e_trans0}), which results in a change of periodic boundary conditions for $H_\textbf{A}$ to antiperiodic ones for $\bar{H}_\textbf{A}$.  This flux is the direct consequence of the Berry phase in graphene. In the limit $W\gg L$ one finds $\bar{\bb{A}} \approx (0, B a^2 \bar{x})$, which corresponds to a constant perpendicular magnetic field in the Landau gauge.

Let us now make a generalized gauge transformation,
\be
\label{e_trans1}
e^{\sigma_z\chi(r) -i\varphi(\theta)}\bar{H}_\textbf{A}e^{\sigma_z\chi(r) +i\varphi(\theta)} =\bar{H}_0, \qquad \bar{H}_0= v\, \bb{\sigma} \bar{\bb{p}},
\e
where we introduced the functions
\be
\label{e_trans2}
\chi(r)= -\frac{e B}{4 c \hbar}(r^2-a^2)+\gamma\, \ln r/a, \qquad \varphi(\theta) = \gamma\,\theta, \qquad \gamma= \frac{e B}{4 c \hbar}\frac{R^2-a^2}{\ln R/a}.
\e
The transformation (\ref{e_trans1},\ref{e_trans2}) is completely analogous to the one introduced in Eqs.~(1,2) of the main text. The function $\chi(r)$ is chosen such that $\chi(a)=\chi(R)=0$ hence the boundary conditions at the leads are unaffected by the transformation (\ref{e_trans1}). Still, the boundary condition in $\theta$ has to change from antiperiodic, for the Hamiltonian $\bar{H}_\textbf{A}$, to quasi-periodic, for $\bar{H}_0$, since $\varphi(2\pi) = 2\pi \gamma$. Thus the Hamiltonian $\bar{H}_0$ corresponds to a graphene cylinder pierced by a magnetic flux $\bar{\Phi}= \Phi /2 L + \Phi_0/2$, where $\Phi= \pi (R^2-a^2) B$ is nothing but the magnetic flux through the Corbino disk, $L=\ln R/a$,  and $\Phi_0=hc/e$ is the flux quantum. 

Thus, the Dirac-point conductance of a ballistic graphene sample in Corbino-disk geometry is equivalent to that of a graphene cylinder pierced by the magnetic flux $\bar{\Phi}$. If the leads are modeled by infinitely doped graphene we recover the well-known result for the conductance  \cite{e_Tworzydlo06},
\be
\label{e_ballG}
\textrm{G}  = \frac{4e^2}{h} \s_{n=-\infty}^\infty \frac{1}{\cosh^2 (n-\Phi/\Phi_c+1/2)L }, \qquad \Phi_c=2\Phi_0 \ln R/a,\qquad \Phi=\pi(R^2-a^2)B.
\e
The result (\ref{e_ballG}) has been obtained directly by wave matching in Refs. \cite{e_katsnelson} and \cite{e_rycerz}. Using Poisson summation, one can rewrite the ballistic conductance (\ref{e_ballG}) in the equivalent form
\be
\label{e_ballG2}
\textrm{G}=\frac{4e^2}{h}\frac{2}{L}\lt(1-\frac{2\pi^2}{L} \s_{n=0}^\infty \frac{1+\cos (2\pi\Phi/\Phi_c) \cosh (\pi^2(2n+1)/L)}{(\cos(2\pi\Phi/\Phi_c)+\cosh(\pi^2(2n+1)/L))^2} \rt).
\e
Restricting ourselves to realistic values of the parameter $L=\ln R/a$ in the Corbino disk, $L\ll \pi^2$, we find \cite{e_katsnelson}
\be
\textrm{G} = \frac{4e^2}{h} \frac{2}{L}\lt(1-\frac{4\pi^2}{L}e^{-\pi^2/L} \cos\lt( \frac{2\pi\Phi}{\Phi_c}\rt)\rt), \qquad L =\ln \frac{R}{a} \ll \pi^2.
\e
Thus, the transformations (\ref{e_trans0},\ref{e_trans1}) are proven to be useful in the analysis of the ballistic conductance at the Dirac point. These transformations can also be applied more widely to relate the Dirac-point conductance of graphene with point-like impurities in Corbino-disk and cylinder geometries. 

\subsection{Green's functions in Corbino geometry}

In order to apply the transformations (\ref{e_trans1},\ref{e_trans2}) to a sample with impurities we have to consider the Keldysh Green's function at zero energy (the Dirac point). In order to study transport quantities we introduce the dimensionless counting fields $\zeta_\pm$ at the interfaces between the leads and the sample. This requires an extension of the Green's function into the Keldysh space. The zero-energy Green's function for the Corbino disk in the absence of impurities satisfies the following equation
\be
\label{e_Kproblem1}
\bpm i0-H_\textbf{A}-\mu(r) & -\hbar v_{\bb{n}} \zeta_+\,\delta(r-a) \\ -\hbar v_{\bb{n}}\zeta_-\,\delta(r-R) & -i0 -H_\textbf{A} -\mu(r) \epm G_{\textbf{A}}(\bb{r},\bb{r}')=\delta(\bb{r}-\bb{r}'),\qquad v_{\bb{n}}=v (\bb{\sigma}\cdot\bb{r})/r,
\e
where the Green's function $G_{\textbf{A}}(r,\theta;r',\theta')$ has to be periodic with respect to the variables $\theta$ and $\theta'$. The model of infinitely doped leads \cite{e_Tworzydlo06} formally corresponds to the choice $\mu(r<a)=\mu(r>R)= - \infty$, $\mu(a\leq r \leq R)=0$. The conductance and other transport quantities are, then, expressed as the derivatives with respect to the counting fields from the corresponding statistical sum as discussed in Section \ref{e_sec:counting}. 

Using the transformations (\ref{e_trans1},\ref{e_trans2}) we relate the Green's function $G_{\textbf{A}}$ associated with the Hamiltonian $H_{\textbf{A}}$ in Corbino geometry to the Green's function $\bar{G}_0$ associated with the Hamiltonian $\bar{H}_0$ for graphene cylinder as
\be
\label{e_Gtrans}
G_{\textbf{A}}(r,\theta;r',\theta') = \frac{1}{\sqrt{r\, r'}}e^{i(\varphi(\theta)-\varphi(\theta'))}\;
e^{\sigma_z(\chi(r)-i\theta/2)} \bar{G}_0(\bar{x},\theta;\bar{x}',\theta')\, e^{\sigma_z (\chi(r')+i\theta'/2)}, \qquad \bar{x}=\ln r/a . 
\e
It is evident from Eq.~(\ref{e_Gtrans}) that the Green's function $\bar{G}_0$ satisfies the quasi-periodic boundary conditions in the variables $\theta$ and $\theta'$ such that $\bar{G}_0(\theta+2\pi;\theta') =- e^{-2\pi i \gamma}  \bar{G}_0(\theta,\theta') $ and  $\bar{G}_0(\theta;\theta'+2\pi) =- e^{2\pi i \gamma} \bar{G}_0(\theta,\theta')$, which corresponds to the cylinder pierced by the magnetic flux $\bar{\Phi}$ (note that $e^{i\pi \sigma_z} \bar{G}_0e^{-i\pi \sigma_z}=-\bar{G}_0$). Using the transformation (\ref{e_Gtrans}) in Eq.~(\ref{e_Kproblem1}) we obtain
\be
\label{e_Kproblem2}
\bpm i0-\bar{H}_0-\bar{\mu}(r) & -\hbar v \sigma_x \zeta_+\,\delta(\bar{x}) \\ -\hbar v \sigma_x\zeta_-\,\delta(\bar{x}-L) & -i0 -\bar{H}_0 -\bar{\mu}(r) \epm \bar{G}_0(\bar{x},\theta;\bar{x}',\theta')=\delta(\bar{x}-\bar{x}')\delta(\theta-\theta'),\qquad \bar{\mu}(r)=r \,\mu(r),
\e
where $L=\ln R/a$. Evidently the infinitely doped leads in Corbino-disk geometry correspond to infinitely doped leads in cylinder geometry, i.e. $\bar{\mu}(r<a)=\bar{\mu}(r>R)= - \infty$ and $\bar{\mu}(a\leq r \leq R)=0$.

The same transformation (\ref{e_Gtrans}) relates the impurity T-matrix in the Corbino disk, $T_\textbf{A}$, to that in the cylinder geometry, $\bar{T}$. Restricting ourselves to $s$-wave scattering we find
\be
\bar{T}(\bar{x},\theta) = \frac{1}{r} e^{\sigma_z(\chi(r)+i\theta/2)}T_\textbf{A}(\bb{r}) e^{\sigma_z(\chi(r)-i\theta/2)},
\e
where $\bb{r}$ is the impurity coordinate, which is parameterized by $\bar{x}=\ln r/a$ and the angle $\theta$. Thus, the results of the main text also apply to the Corbino-disk geometry.


\begin{thebibliography}{99}

\bibitem{geim07}  
K.\,S.\,Novoselov, A.\,K. Geim, S.\,V. Morozov, D.\,Jiang, Y.\,Zhang, S.\,V. Dubonos, I.\,V.  Grigorieva,
and A.\,A. Firsov, Science {\bf 306}, 666 (2004);
K.\,S.\,Novoselov, D.\,Jiang, T.\,Booth, V.\,V. Khotkevich,
S.\,M.\,Morozov, and A.\,K.\,Geim, PNAS  {\bf 102}, 10451 (2005);
A.\,K.\,Geim and K.\,S.\,Novoselov, Nature Materials {\bf 6},
183-191 (2007);
A.\,K.\,Geim, Science \textbf{324}, 1530 (2009);
K.\,S.\,Novoselov, Rev.\ Mod.\ Phys. \textbf{83}, 837 (2011); 
A.\,K.\,Geim, Rev.\ Mod.\ Phys. \textbf{83}, 851 (2011).

\bibitem{graphene-review}  A.\,H.\,Castro Neto, F.\,Guinea, N.\,M.\,R. Peres,
K.\,S.\,Novoselov, and A.\,K.\,Geim, Rev.\ Mod.\ Phys. {\bf 81}, 109 (2009).

\bibitem{Elias09}
D.\,C.\,Elias, R.\,R.\,Nair, T.\,M.\,G.\,Mohiuddin, S.\,V.\,Morozov, P.\,Blake, M.\,P.\,Halsall,
A.\,C.\,Ferrari, D.\,W. Boukhvalov, M.\,I.\,Katsnelson, A.\,K.\,Geim, and K.\,S.\,Novoselov, Science \textbf{323}, 610 (2009).

\bibitem{ryu08} S.\,Ryu, M.\,Y.\,Han, J.\,Maultzsch, T.\,F. Heinz, P.\,Kim, M.\,L.\,Steigerwald and L.\,E.\,Brus, Nano Lett. {\bf 8}, 4597 (2008).

\bibitem{burgess11} B.\,R. Matis, F.\,A. Bulat, A.\,L.\,Friedman, B.\,H.\,Houston, and J.\,W.\,Baldwin, Phys.\ Rev.\ B {\bf 85}, 195437 (2012);
J.\,S.\,Burgess, B.\,R.\,Matis, J.\,T.\,Robinson, F.\,A.\,Bulat, F.\,K.\,Perkins, B.\,H.\,Houston, and J.\,W.\,Baldwin, Carbon {\bf 49}, 4420 (2011);
B.\,R.\,Matis, J.\,S.\,Burgess, F.\,A.\,Bulat, A.\,L.\,Friedman, B.\,H.\,Houston, and J.\,W.\,Baldwin, ACS Nano {\bf 6}, 17 (2012). 

\bibitem{Katoch10}
J.\,Katoch, J.-H.\,Chen, R.\,Tsuchikawa, C.\,W.\,Smith, E.\,R.\,Mucciolo, and M.\,Ishigami, Phys.\ Rev.\ B \textbf{82}, 081417(R) (2010).

\bibitem{robinson10}
J.\,T.\,Robinson, J.\,S.\,Burgess, C.\,E.\,Junkermeier, S.\,C.\,Badescu, T.\,L.\,Reinecke, F.\,K.\,Perkins, M.\,K.\,Zalalutdniov, J.\,W.\,Baldwin, J.\,C.\,Culbertson, P.\,E.\,Sheehan, and E.\,S.\,Snow,
Nano Lett. {\bf 10}, 3001 (2010).

\bibitem{schedin07} F.\,Schedin, A.\,K.\,Geim, S.\,V.\,Morozov, E.\,W.\,Hill, P.\,Blake,
M.\,I.\,Katsnelson, and K.\,S.\,Novoselov, Nature Mater. {\bf 6}, 652 (2007). 

\bibitem{Liu08}
L.\,Liu, S.\,Ryu, M.\,R.\,Tomasik, E.\,Stolyarova, N.\,Jung, M.\,S.\,Hybertsen, M.\,L.\,Steigerwald, L.\,E.\,Brus, and G.\,W.\,Flynn, \textit{et al.}, Nano Lett. \textbf{8}, 1965 (2008).

\bibitem{fuhrer09}
J.-H.\,Chen, W.\,G.\,Cullen, C.\,Jang, M.\,S.\,Fuhrer, E.\,D.\,Williams,
Phys.\ Rev.\ Lett. {\bf 102}, 236805 (2009).

\bibitem{fuhrer11}
J.-H.\,Chen, W.\,G.\,Cullen, E.\,D.\,Williams, and M.\,S.\,Fuhrer,
Nature Phys. {\bf 7}, 535 (2011).

\bibitem{balandin} D.\,Teweldebrhan and A.\,A.\,Balandin, Appl. Phys. Lett. {\bf 94}, 013101 (2009); {\bf 95}, 246102 (2009); G.\,Liu, D.\,Teweldebrhan, and A.\,A.\,Balandin, IEEE Transactions on Nanotechnology {\bf 10}, 865 (2011). 

\bibitem{Bouchiat}
B.\,M.\,Kessler, C.\,\"O.\,Girit, A.\,Zettl, and V.\,Bouchiat,  Phys.\ Rev.\ Lett. {\bf 104}, 047001 (2010). 

\bibitem{Lichtenstein}
T.\,O.\,Wehling \textit{et al.}, Phys.\ Rev.\ B \textbf{75}, 125425 (2007); {\it ibid}, B {\bf 80}, 085428 (2009); 
T.\,O.\,Wehling, S.\,Yuan, A.\,I.\,Lichtenstein, A.\,K.\,Geim, M.\,I.\,Katsnelson, Phys.\ Rev.\ Lett.\ \textbf{105}, 056802 (2010).

\bibitem{andrei08}
X.\,Du, I.\,Skachko, A.\,Barker, and E.\,Y.\,Andrei, Nature Nanotechnology \textbf{3}, 491 (2008);
X.\,Du, I.\,Skachko, and E.\,Y.\,Andrei, Int. J. of Mod. Phys. B \textbf{22}, 4579 (2008);
D.\,A.\,Abanin, I.\,Skachko, X.\,Du, E.\,Y.\,Andrei, and L.\,S.\,Levitov, Phys.\ Rev.\ B \textbf{81}, 115410 (2010).

\bibitem{bolotin}
K.\,I.\,Bolotin, K.\,J.\,Sikes, Z.\,Jiang, M.\,Klima, G.\,Fudenberg, J.\,Hone, P.\,Kim, and H.\,L.\,Stormer, Solid State Communications \textbf{146}, 351 (2008); K.\,I.\ Bolotin, K.\,J.\,Sikes, J.\,Hone, H.\,L.\ Stormer and P.\,Kim, Phys.\ Rev.\ Lett. {\bf 101}, 096802 (2008).

\bibitem{ostrovsky06} P.\,M.\,Ostrovsky, I.\,V.\,Gornyi, and A.\,D.\,Mirlin, Phys.\ Rev.\ B \textbf{74}, 235443 (2006).

\bibitem{Stauber07} T.\,Stauber, N.\,M.\,R.\,Peres, and F.\,Guinea,  Phys.\ Rev.\ B {\bf 76}, 205423 (2007).

\bibitem{novikov08} D.\,S.\,Novikov, Phys.\ Rev.\ B \textbf{76}, 245435 (2007).

\bibitem{basko08} D.\,M.\,Basko, Phys.\ Rev.\ B \textbf{78}, 115432 (2008).

\bibitem{robinson08} J.\,P.\,Robinson, H.\,Schomerus, L.\,Oroszlany, and
V.\,I.\,Falko, Phys.\ Rev.\ Lett. \textbf{101}, 196803 (2008).

\bibitem{pereira08} V.\,M.\,Pereira, J.\,M.\,B.\,Lopes dos Santos, and A.\,H.\,Castro
Neto, Phys.\ Rev.\ B \textbf{77}, 115109 (2008).

\bibitem{Yuan10}
S.\,Yuan, H.\,De\,Raedt, and M.\,I.\,Katsnelson, Phys. Rev. B {\bf 82}, 115448
(2010).


\bibitem{novoselov05b}  K.\,S.\,Novoselov, A.\,K.\,Geim, S.\,V.\,Morozov,
D.\,Jiang, M.\,I.\,Katsnelson, I.\,V.\,Grigorieva, S.\,V.\,Dubonos,
and A.\,A.\,Firsov,  Nature {\bf 438}, 197 (2005).

\bibitem{kim05}  Y.\,Zhang, Y.-W.\,Tan, H.\,L.\,Stormer,
and P.\,Kim, Nature {\bf 438}, 201 (2005);
Y.-W.\,Tan, Y.\,Zhang,  H.\,L.\,Stormer,
and P.\,Kim, Eur. Phys. J. Special Topics {\bf 148}, 15 (2007).

\bibitem{evers08}   F.\,Evers and A.\,D.\,Mirlin, Rev.\ Mod.\ Phys. \textbf{80}, 1355 (2008).

\bibitem{ostrovsky07}  P.\,M.\,Ostrovsky, I.\,V.\,Gornyi, and A.\,D.\,Mirlin, Phys.\ Rev.\ Lett. \textbf{98}, 256801 (2007);
Eur.\ Phys.\ J. Special Topics \textbf{148}, 63 (2007).

\bibitem{ponomarenko11} L. A. Ponomarenko,  A. K. Geim, A. A.
Zhukov, R. Jalil, S. V. Morozov, K. S. Novoselov, V. V. Cheianov, V. I. Fal'ko,
K. Watanabe, T. Taniguchi, and R. V. Gorbachev, Nature Phys. {\bf 7}, 958
(2011).

\bibitem{novoselov07}
K.\,S.\,Novoselov, Z.\,Jiang, Y.\,Zhang, S.\,V.\,Morozov, H.\,L.\,Stormer, U.\,Zeitler, J.\,C.\,Maan, G.\,S.\,Boebinger, P.\,Kim, and A.\,K.\,Geim, Science {\bf 315}, 1379 (2007).

\bibitem{AQHE} 
K.\,S.\,Novoselov, A.\,K.\,Geim, S.\,V.\,Morosov, D.\,Jiang, M.\,I.\,Katsnelson,
I.\,V.\,Grigorieva, S.\,V.\,Dubonos, and A.\,A.\,Firsov,  Nature \textbf{438}, 197 (2005); Y.\,Zhang, Y.-W.\,Tan, H.\,L.\,Stormer, and P.\,Kim,
Nature \textbf{438}, 201 (2005). 

\bibitem{Abanin07a} 
D.\,A.\,Abanin, K.\,S.\,Novoselov, U.\,Zeitler, P.\,A.\,Lee, A.\,K.\,Geim, and
L.\,S.\,Levitov, Phys.\ Rev.\ Lett. \textbf{98}, 196806 (2007). 

\bibitem{Ong08} 
J.\,G.\,Checkelsky, L.\,Li, and N.\,P.\,Ong, Phys.\,Rev.\,Lett. \textbf{100}, 206801 
(2008); Phys.\ Rev.\ B \textbf{79}, 115434 (2009). 

\bibitem{Kim12} 
Y.\,Zhao, P.\,Cadden-Zimansky, F.\,Ghahari, and P.\,Kim, Phys.\ Rev.\ Lett.
\textbf{108}, 106804 (2012). 

\bibitem{Bolotin09} 
K.\,I.\,Bolotin, F.\,Ghahari, M.\,D.\,Shulman, H.\,L.\,Stormer, and P.\,Kim, Nature
\textbf{462}, 196 (2009). 

\bibitem{Du09} 
X.\,Du, I.\,Shachko, F.\,Duerr, A.\,Luican, and E.\,Y.\,Andrei, Nature \textbf{462},
192 (2009). 

\bibitem{Ostrovsky08}
P.\,M.\,Ostrovsky, I.\,V.\,Gornyi, and A.\,D.\,Mirlin, Phys.\ Rev.\ B \textbf{77},
195430 (2008).

\bibitem{roche-localization}
A. Lherbier, B. Biel, Y.-M. Niquet, and S. Roche,
Phys. Rev. Lett. {\bf 100}, 036803 (2008);
A. Lherbier, Simon M.-M. Dubois, X. Declerck, S. Roche,
Y.-M. Niquet, and J.-C. Charlier,
Phys. Rev. Lett. {\bf 106}, 046803 (2011);
N. Leconte, A. Lherbier, F. Varchon, P. Ordejon, S. Roche, and J.-C.
Charlier, Phys. Rev. B {\bf 84}, 235420 (2011). 

\bibitem{roche-vacancies}
A. Cresti, F. Ortmann, T. Louvet, D. Van Tuan,
and S. Roche, Phys. Rev. Lett. {\bf 110}, 196601 (2013). 

\bibitem{mayou}
G. T. de Laissardiere and D. Mayou, Mod. Phys. Lett. B {\bf 25}, 1019 (2011);
arXiv:1212.3997.

\bibitem{Titov10}
M.\,Titov, P.\,M.\,Ostrovsky, I.\,V.\,Gornyi, A.\,Schuessler, and A.\,D.\,Mirlin, Phys.\ Rev.\ Lett. \textbf{104}, 076802 (2010).  

\bibitem{Ostrovsky10}
P.\,M.\,Ostrovsky, M.\,Titov, S.\,Bera, I.\,V.\,Gornyi, A.\,D.\,Mirlin,
Phys.\ Rev.\ Lett. \textbf{105}, 266803 (2010). 

\bibitem{Schelter11}
J.\,Schelter, P.\,M.\,Ostrovsky, I.\,V.\,Gornyi, B.\,Trauzettel, and M.\,Titov,
Phys.\ Rev.\ Lett. \textbf{106}, 166806 (2011). 

\bibitem{Tworzydlo06}
J.\,Tworzyd{\l}o, B.\,Trauzettel,  M.\,Titov, A.\,Rycerz, and C.\,W.\,J.\,Beenakker,
Phys.\ Rev.\ Lett. \textbf{96}, 246802 (2006).

\bibitem{Schuessler09} A.\,Schuessler, P.\,M.\,Ostrovsky, I.\,V.\,Gornyi, and A.\,D.\,Mirlin, Phys.\ Rev.\ B \textbf{79}, 075405 (2009).

\bibitem{EPAPS} Online Supplementary Material

\bibitem{Ortmann13}
F.\,Ortmann and S.\,Roche, Phys.\ Rev.\ Lett \textbf{110}, 086602 (2013).

\bibitem{nomura08} K. Nomura, S. Ryu, M. Koshino, C. Mudry, and A.
Furusaki, Phys. Rev. Lett. {\bf 100}, 246806 (2008).

\bibitem{Avishai93}
Y.\,Avishai, and M.\,Ya.\,Azbel, and S.\,A.\,Gredeskul,  Phys.\ Rev.\ B \textbf{48}, 17280 (1993); M.\,Ya.\,Azbel, Phys.\ Rev.\ B
\textbf{49}, 5463 (1994);
M.\,Ya.\,Azbel and B.\,I.\,Halperin, Phys.\ Rev.\ B \textbf{52}, 14098 (1995);
S.\,A.\,Gredeskul, M.\,Zusman, Y.\,Avishai, and M.\,Ya.\,Azbel, Phys.\ Rep. {\bf 288}, 223 (1997). 


\end{thebibliography}

\begin{thebibliography}{99}

\bibitem{e_Tworzydlo06}
J.~Tworzyd{\l}o, B.~Trauzettel,  M.~Titov, A.~Rycerz, and C.~W.~J.~Beenakker, Phys.\ Rev.\ Lett. \textbf{96}, 246802 (2006).

\bibitem{e_Titov10}
M.~Titov, P.~M.~Ostrovsky, I.~V.~Gornyi, A.~Schuessler, and A.~D.~Mirlin, Phys.\
Rev.\ Lett. \textbf{104}, 076802 (2010).  

\bibitem{e_Ostrovsky10}
P. M. Ostrovsky, M. Titov, S. Bera, I. V. Gornyi, A. D. Mirlin,
Phys.\ Rev.\ Lett. \textbf{105}, 266803 (2010). 

\bibitem{e_ostrovsky07}  P.M. Ostrovsky, I.V. Gornyi, and A.D. Mirlin,
  Phys. Rev. Lett. \textbf{98}, 256801 (2007);
Eur. Phys. J. Special Topics \textbf{148}, 63 (2007).

\bibitem{e_wimmer}
A.\,Rycerz, P.\,Recher, M.\,Wimmer,
Phys.\ Rev.\ B \textbf{80}, 125417 (2009).

\bibitem{e_katsnelson}
M.\,I.\,Katsnelson, Europhys.\ Lett. \textbf{89}, 17001 (2010). 

\bibitem{e_rycerz}
A.\,Rycerz, Phys.\ Rev. B \textbf{81}, 121404(R) (2010) 
\end{thebibliography}
\end{document}